\documentclass[amsmath,amssymb,twocolumn,superscriptaddress]{revtex4}
\usepackage{graphicx}% Include figure files
\usepackage{dcolumn}% Align table columns on decimal point

\def\be{\begin{equation}}
\def\ee{\end{equation}}
\def\beq{\begin{eqnarray}}
\def\eeq{\end{eqnarray}}

\usepackage{bm}% bold math
\usepackage{graphicx}
\usepackage{dcolumn}
\usepackage{amsmath,enumitem}
\usepackage[utf8]{inputenc}
\usepackage{graphicx, psfrag}
\usepackage{amssymb}
\usepackage[colorlinks=true, citecolor=blue, urlcolor = blue, linkcolor= red, bookmarks=true]{hyperref}
\usepackage{float}
\usepackage{amsmath}
\usepackage{tensor}
\usepackage{amsfonts}
\usepackage{dcolumn}
\usepackage{hyperref}
\usepackage{subfigure}
\usepackage{pgfplots}
\usepackage{epstopdf}
\usepackage{booktabs}
\usepackage{amsmath, amssymb}
\usepackage{multirow}
\usepackage{graphicx}
\usepackage[usenames,dvipsnames]{xcolor}% for marking revised text in blue
\begin{document}

\title{Lorentz Symmetry Breaking Traversable Wormhole Models Supported by Einasto Dark Matter}

\author{Rana Muhammad Zulqarnain }
\email{ranazulqarnain7777@gmail.com}
\affiliation{School of Business, Xian International University, Xian, 710077, Shaanxi, China.}

\author{M. Yousaf}
\email{myousaf.math@gmail.com}
\affiliation{Department of Mathematics, Virtual University of Pakistan,\\ 54-Lawrence Road, Lahore 54000, Pakistan.}
\affiliation{Research Center of Astrophysics and Cosmology, Khazar University,
Baku, AZ1096, 41 Mehseti Street, Azerbaijan.}

\author{Allah Ditta}
\email{mradshahid01@gmail.com}
\affiliation{School of Science, Walailak University, Nakhon Si Thammarat, 80160, Thailand}

\author{Asifa Ashraf}
\email{asifamustafa3828@gmail.com}
\affiliation{School of Mathematical Sciences, Zhejiang Normal University, Jinhua, Zhejiang 321004, China}

\author{Ahmadjon Abdujabbarov}
\email{ahmadjonab@gmail.com}
\affiliation{School of Physics, Harbin Institute of Technology, Harbin 150001, People's Republic of China}

\author{Farruh~Atamurotov}
\email{atamurotov@yahoo.com}
\affiliation{Kimyo International University in Tashkent, Shota Rustaveli str. 156, Tashkent 100121, Uzbekistan}

\pacs{04.40.-b; 04.40.Dg; 04.50.Kd; 04.25.Nx.}

\begin{abstract}
In this work, we investigate static and spherically symmetric traversable wormhole configurations in Kalb Ramond gravity, where spontaneous Lorentz symmetry breaking is induced by a non vanishing vacuum expectation value of the antisymmetric Kalb Ramond field. The matter sector is modeled by the Einasto dark matter density profile, which enables us to derive an analytical shape function involving the incomplete Gamma function. The resulting geometry is examined through the standard throat, flare-out, and asymptotic requirements, as well as the corresponding embedding diagrams provide a geometric visualization of the wormhole structure. The energy density remains positive in the considered domain, whereas the radial null energy condition is violated in the vicinity of the throat, indicating that the exotic matter required for traversability can be localized within a restricted region. The internal structure is further characterized through the complexity factor, which is most pronounced near the throat and gradually approaches zero at larger radial distances. We also investigate the total gravitational energy, active gravitational mass, average pressure, and equilibrium behavior through the TOV equation and pressure anisotropy. For the constant redshift configuration, the equilibrium is governed by the counterbalancing hydrostatic and anisotropic forces. In addition, the optical properties of the spacetime are explored through the photon sphere, critical impact parameter, light deflection angle, and echo time. Finally, the volume integral quantifier is employed to estimate the total amount of null energy condition violating matter, demonstrating that the exotic contribution can remain concentrated around the wormhole throat. The analysis highlights the combined role of the Kalb Ramond gravitational parameter and the Einasto matter distribution in determining the geometric, energetic, and observational characteristics of the resulting wormhole configurations.
\\\\
\textbf{Keywords:} {Kalb Ramond gravity; Einasto dark matter profile; Traversable wormhole; Lorentz symmetry breaking; Optical observables.}
\end{abstract}
\maketitle

\section{Introduction}

Newton's classical theory interprets gravity as an instantaneous attractive force acting between masses within an absolute, immutable background of space and time. Einstein later revolutionized this viewpoint by proposing that gravitation arises instead from the curvature of spacetime produced by mass energy. According to this relativistic description, space and time merge into a unified four-dimensional continuum whose geometry governs the motion of matter. Einstein's special theory of relativity (ESR) demonstrated that measurements of space and time depend on the observer's state of motion. Later on Einstein's general theory of relativity revealed that massive bodies warp spacetime, thereby altering the flow of time and influencing nearby trajectories and this reformulation successfully accounted for several phenomena that Newtonian theory could not explain, like the anomalous precession of Mercury's orbit and the bending of light by gravity, thereby offering a more complete theoretical scenario for describing the universe.

At the dawn of the twentieth century, Einstein questioned the Newtonian assumption that space and time possessed absolute character and he maintained that a physical theory should not depend on such constructs. The null result of the Michelson Morley experiment, which found no evidence for Earth's motion through the hypothetical luminiferous ether, critically shaped his reasoning. Motivated by this outcome, Einstein formulated a special theory, which is based on a non-accelerated background, which is also known as ESR, resolving longstanding inconsistencies between classical kinematics and Maxwell's electromagnetic theory \cite{einstein1982created}. While Maxwell's prediction of a constant light speed conflicted with the Galilean law of velocity addition, leading Einstein to reexamine how distant clocks are synchronized. Although informed by the Michelson-Morley results, Einstein's development of ESR explicitly avoided introducing dynamical assumptions about motion through the ether \cite{holton1960origins}.

Within ESR, the word special refers to its restriction to inertial reference frames, in which bodies move without external forces, while this theory gave rise to profound consequences such as time dilation, length contraction, and relativistic mass-energy relations. Shortly thereafter, Minkowski reformulated physics within a four-dimensional spacetime, a pivotal conceptual step that provided the mathematical foundation for Einstein's general theory of relativity. The Einstein's general theory of relativity generalizes the relativistic framework to accelerated frames and replaces Newton's gravitational force with a geometric description of spacetime curvature. The theory rests on two central principles \cite{einstein1916foundation}: the principle of relativity, which states that all observers experience the same fundamental laws of physics, and the principle of general covariance, which requires that physical laws retain their form under arbitrary coordinate transformations. The Einstein's general theory of relativity also establishes the equivalence between inertial and gravitational effects, asserting that gravitational and accelerative influences cannot be locally distinguished. Einstein's field equations (EFEs) form the core of Einstein's general theory of relativity, whereas these are ten coupled, nonlinear partial differential equations possessing both hyperbolic and elliptic features. They relate the distribution of matter and energy to the curvature of spacetime through key mathematical objects, including the metric tensor, the Ricci tensor, and various curvature invariants. Together, these equations provide the fundamental framework that governs gravitational dynamics in relativistic settings. The EFEs formalize the manner in which matter, energy, and momentum influence the geometry of spacetime, thereby giving rise to what is perceived as gravitational attraction, while this geometric interpretation enables the EFEs to account for several previously puzzling phenomena, such as the existence of gravitational waves, the dynamics and singular structures of black holes, and the accelerated cosmic expansion. In essence, the EFEs establish the fundamental correspondence between the distribution of matter and the resulting curvature of spacetime. The conceptual development of WHs within the framework of Einstein's general theory of relativity dates back to 1935; however, the term WH itself was introduced later by Wheeler in 1957 \cite{shinkai2015wormhole}. A WH is typically described as a theoretical spacetime configuration capable of linking two separate asymptotically flat regions or remote spatial domains \cite{morris1988wormholes}. Although WHs are defined through their global topological features, their local description relies on identifying a compact two-dimensional spatial surface on an achronal hypersurface, commonly called the throat.

A classical example associated with WH geometry arises from the fully extended Schwarzschild solution. Einstein and Rosen demonstrated that, through an appropriate redefinition of the radial coordinate, the Schwarzschild spacetime can be interpreted as forming a bridge between two exterior regions, now termed the Einstein-Rosenstein bridge \cite{einstein1935particle}. The location of the throat in this interpretation corresponds to the bifurcation of the two spheres of the event horizon and appears as a coordinate singularity within the original Schwarzschild coordinates. Consequently, in its standard formulation, the Einstein-Rosen bridge resembles a static WH but lacks a genuine traversable throat. Although alternative coordinate frameworks can momentarily exhibit a throat-like structure, Fuller and Wheeler later showed that such a configuration is transient in nature, as represented in the Penrose diagram \cite{wheeler1955geons}. Hence, while the maximally extended Schwarzschild geometry admits a dynamic WH interpretation, the static Einstein-Rosen bridge does not meet the criteria for a physically realizable static WH. Consequently, WHs are one of the special solutions of the gravitational field equations of Einstein's general theory of relativity \cite{hawking1975large}, which are theoretical tunnels linking disparate regions of spacetime and constitute a captivating subject in theoretical physics \cite{visser1995lorentzian}. Despite considerable scientific interest, a viable mechanism for their natural formation remains elusive. The principal obstacle stems from the necessity of exotic matter, which exhibits a negative energy density and is not known to exist in macroscopic quantities. Although quantum field theoretic effects permit localized negative energy densities enabling the study of microscopic WHs as investigated by Callan and Maldacena \cite{callan1998brane} in this study they employed the abelian Born-Infeld action to investigate D-brane structure and dynamics through world volume gauge fields and transverse scalar excitations, and it demonstrated that point charges and vortices become BPS saturated when transverse modes excited. The Coulomb point charge solutions effectively model fundamental strings attached to D-branes, while magnetic counterparts describe D1-D3 brane systems, additionally, the S type matrix analysis of small perturbations supports Polchinski's open string boundary conditions, providing deeper insights into D-brane dynamics. This framework is inapplicable to large scale astrophysical WHs, whose characteristics are profoundly distinct \cite{visser1995lorentzian}. The existence of stable, traversable Lorentzian WHs would profound implications, potentially enabling the construction of closed timelike curves (time machines) or shortcuts for interstellar travel, as a result, such structures exist on astrophysical scales, they could give rise to a suite of unique and observable phenomena \cite{ellis1973ether, bronnikov1973scalar, clement1981einstein, hochberg1998dynamic, simonetti2021sensitive, dai2018new}.

This pursuit is further refined into the search for traversable WHs structures of sufficient scale to permit human passage without destructive tidal forces, while first conceptualized in their modern form by Morris and Thorne \cite{morris1988wormholes}.
In their work they presented an accessible discussion of rapid interstellar travel via spacetime WHs, inspired by Carl Sagan's novel Contact which argued that black holes or Schwarzschild WHs unsuitable for traversal and introduces new Einstein field equation solutions describing traversable WHs without horizons. These required exotic matter with radial tension $\tau_0$ exceeding its energy density $\sigma c^2$, violating classical energy conditions although no known material satisfies $\tau_0 > \sigma c^2$, quantum field theory suggests the possible existence of such exotic matter. A particularly profound implication of traversable WHs is their potential to function as time machines, while this arises from inducing a time shift between the WH's two mouths, achievable through methods such as the twin paradox of special relativity or the gravitational redshift of Einstein's general theory of relativity. In such a configuration, closed timelike curves paths that loop back into their own past can form beyond a specific boundary known as the chronology horizon, a Cauchy horizon confined within a compact region \cite{kim1991vacuum}. It is a well established result in Einstein's general theory of relativity that the formation of a WH from Einstein's equations requires a violation of the null energy constraints in the vicinity of the throat, while this conclusion is rigorously proven for static configurations. For dynamic WHs, the definition of a throat must be generalized, as demonstrated by Hochberg and Visser \cite{hochberg1998dynamic}, with alternative formulations also being proposed \cite{hayward1999dynamic}. Conventionally, from an external vantage point, a WH entrance is presumed to resemble a localized gravitational source, analogous to a black hole or a compact star \cite{hawking1988wormholes,singh2020conformally,mustafa2021traversable,hassan2021traversable,de2023epicyclic,mustafa2023imprints,yousaf2023cylindrical,dai2019observing}.

Beyond such localized objects, the Universe may also contain structures infinite in spatial extent, such as cosmic strings, while star like and black hole geometries are naturally described by spherical symmetry, the simplest models for string like objects are founded upon cylindrical symmetry. The feasibility of constructing WH models within such a cylindrically symmetric framework recently investigated \cite{bronnikov2009cylindrical,de2020general}. Dai \emph{et al.} \cite{dai2020form} studied a simple model for WH formation by placing two massive objects on parallel branes representing separate universes, while gravitational attraction between the objects competes with brane tension, and when the attraction dominates, the branes deform and a WH forms. Their analysis showed that more massive and compact objects favor this process, suggesting that WHs are most likely to occur near black holes, compact stars, or massive relics and the proposed mechanism also applies to WHs connecting objects within the same universe. These studies reveal that the necessary conditions for the existence of static WH solutions differ significantly from their spherically symmetric counterparts, nevertheless, developing realistic cylindrically symmetric WH models remains a formidable challenge. Although numerous solutions derived, none exhibit two flat asymptotic regions or string like asymptotics characterized by flat spacetime with an angular deficit which is a prerequisite for the WH entrances to manifest as localized objects in our Universe, furthermore, it showed that a static, cylindrically symmetric WH that is asymptotically flat at both ends inevitably requires matter sources with negative energy density \cite{bronnikov2009cylindrical,di2017spin}. A dominant yet enigmatic constituent of the Universe is dark matter, which is detected not by its emission of light but through its gravitational influence, while luminous matter accounts for a mere $\sim 3\%$ of the Universe's content, with the remainder consisting of DE as well as dark matter, the latter comprising approximately $27\%$ of the total energy density \cite{salucci2019distribution}. The existence of dark matter was first inferred by Zwicky \cite{zwicky1979masses} through the application of the virial theorem to galactic clusters, while Brinks and Klein \cite{brinks1988dark} studied high-resolution H-I observations of the blue compact dwarf galaxy II Zwicky 40 which revealed two interacting kinematical systems composed of neutral hydrogen clouds and the optical emission lied at the center of the northern cloud, likely due to interaction induced star formation. Each cloud had a total mass an order of magnitude greater than its visible mass, indicating a significant dark matter component within II Zwicky 40. Further, evidence arises from the universal rotation curves of spiral galaxies, which illustrate the gravitational footprint of dark matter halos \cite{roberts1978twenty,roszkowski2018wimp,dai2020form,de2021epicyclic}.

Leveraging such observational data, including the Navarro-Frenk-White profile and flat rotation curves, Rahaman \emph{et al.} \cite{rahaman2014possible} demonstrated that the halos of galaxies could potentially contain traversable WHs. This seminal work stimulated significant research into this phenomenon within various gravitational frameworks \cite{sharif2016wormhole}, subsequent investigations explored WH solutions in isothermal galactic halos using universal rotation curves and Navarro-Frenk-White profiles \cite{rahaman2016study}, the viability of traversable WHs in the Dragonfly 44 galaxy (concluding they are only possible with a generalized Navarro-Frenk-White model) \cite{islam2019formation}, and the formation of spherically symmetric WHs in halos with isotropic pressure \cite{xu2020possibility}. More recent studies extended this line of inquiry to WHs with topological defects in dark matter halos \cite{ovgun2021evolving}, while the gravitational lensing signatures of such galactic WHs also examined \cite{kuhfittig2014gravitational}. The concept of complexity for self-gravitating anisotropic systems formalized by Herrera \cite{herrera2018new}, who derived a corresponding complexity factor by obtaining expressions for a generic mass function and applying the Tolman mass formalism alongside structural scalars from a decomposition of the Riemann curvature tensor. In related work, Herrera \emph{et al.} \cite{herrera2009dynamics} analyzed the fundamental dynamics of gravitational collapse, utilizing the Israel-Stewart formalism to describe viscous dissipation incorporating bulk and shear viscosity. Herrera also provided solutions for the relativistic gravitational collapse of systems with dissipative energy \cite{herrera2002relativistic}. Further expanding on this, Herrera and Santos \cite{herrera2004dynamics} investigated dissipative gravitational collapse, accounting for heat transport and free streaming radiation within the Misner-Sharp approach, complementary to these studies, Yousaf and collaborators have extensively analyzed the complexity factor for various astrophysical systems in the context of modified gravity theories \cite{bhatti2021electromagnetic,yousaf2021quasi}.

The present study is devoted to the construction and analysis of static and spherically symmetric traversable WH configurations in scenario of our considered gravitational Kalb Ramond theory. In particular, we consider a Morris-Thorne geometry along with a spontaneously Lorentz symmetry breaking Kalb Ramond scenario and model the matter sector through the Einasto dark matter density profile, however by combining the modified gravitational field equations with the Einasto distribution.
The Einasto dark matter density profile is one of several analytical families proposed in \cite{einasto1969andromeda} was initially purposed to demonstrate distinguishes between two types of dark matter: one explaining the high rotation speeds of spiral galaxies and another preventing the Universe from expanding indefinitely. While the cold dark matter model successfully accounts for large scale cosmic structures, it fails to match observations of dwarf galaxies, which are dominated by dark matter at kiloparsec scales. Therefore, the assumption that dark matter is mainly cold appears difficult to maintain, while the study investigated how mass resolution and force softening influence the density profiles of cold dark matter halos in cosmological N-body simulations presented in \cite{moore1998resolving}. However, higher resolution reveals dense progenitor halos and over 1,000 surviving subhalos within a single virialized system and the inner density profiles become steeper, following $\sigma(r) \sim r^{-1.4}$, though convergence is uncertain, such steep profiles may challenge the cold dark matter model, as they could conflict with observed galactic halo kinematics and gravitational arc properties. Whereas, Einasto density distribution was applied to characterize the M31 galaxy \cite{einasto1969andromeda} which subsequently extended to describe a range of nearby galaxies as well as the Milky Way too \cite{einasto1972galactic}. These galactic models are composed of multiple components, each representing a distinct, physically homogeneous stellar population defined by unique parameters. Owing to its empirical success, the Einasto profile presents a promising foundation for constructing physically motivated WH solutions.  We derive an analytical expression for the WH shape function as well as investigate the corresponding geometrical and physical viability conditions. The role of the Kalb Ramond gravitational parameter is examined together with the characteristic parameters of the Einasto profile in determining the throat geometry, energy condition behavior, internal structural complexity, equilibrium properties, and optical characteristics of the resulting wormhole (WH) spacetime. This scenario therefore provides a unified setting for exploring how Lorentz symmetry breaking gravitational effects and a realistic dark matter distribution can jointly influence the existence and physical properties of traversable WH configurations.
We follow thge following structure for this manuscript; the theoretical scenario of Kalb Ramond gravity in section~II, while in section~III, we formulate the modified field equations for a static and spherically symmetric WH geometry, however section~IV discuss the construction of the WH shape function with Einasto dark matter density profile and to the examination of the fundamental geometrical requirements for traversability, whereas In section~V, we investigate the corresponding embedding geometry, while section~VI discusses the physical properties of the matter distribution together with the relevant energy conditions. The complexity factor is analyzed in section~VII and in section~VIII, we examine the total gravitational energy, active gravitational mass, as well as average pressure of the resulting configurations, while Section~IX presents the equilibrium analysis based on the generalized TOV equation and pressure anisotropy. In section~X deals with optical properties of the WH spacetime, including the photon sphere, light-deflection angle, echo time, and the volume integral quantifier used to estimate the amount of exotic matter, whereas the main results are given in section~XI.

\section{THEORETICAL FORMULATION OF NON-MINIMALLY COUPLED Kalb Ramond GRAVITY WITH SPONTANEOUS LORENTZ SYMMETRY BREAKING}

In this part of our manuscript, we will discuss the dynamical features of the Kalb Ramond model, with particular emphasis on the choice of its vacuum expectation value and its role in the gravitational sector. According to the literature, the Kalb Ramond field arises in bosonic string theory as a rank-two antisymmetric tensor field $\mathcal{B}_{\eta\aleph}$~\cite{kalb1974classical}. Using the differential-form formalism, this antisymmetric field can be expressed in terms of potential two-form;
\[
\mathcal{B}_2=\frac{1}{2}\mathcal{B}_{\eta\aleph}dx^\eta\wedge dx^\aleph ,
\]
whose associated field strength is given by
\[
\mathcal{H}_3=d\mathcal{B}_2, ~\mathcal{H}_{\lambda\eta\aleph}\equiv \partial_{[\lambda}\mathcal{B}_{\eta\aleph]} .
\]
The Kalb Ramond potential also possesses the gauge freedom
\[
\mathcal{B}_2\rightarrow \mathcal{B}_2+d\Lambda_1,
\]
where $\Lambda_1$ denotes an arbitrary one-form.

In certain theoretical settings, the self-interaction of the Kalb Ramond field may generate non-trivial vacuum configurations capable of breaking both symmetries like gauge as well as Lorentz~\cite{kostelecky2004gravity}. More specifically, one can say the self-interaction of the antisymmetric field can be characterized through a potential written as
\[
V=V\left(\mathcal{B}_{\eta\aleph}\mathcal{B}^{\eta\aleph}\pm b_{\eta\aleph}b^{\eta\aleph}\right)
\]
can produce a non-vanishing vacuum expectation value satisfying $\mathcal{B}_{\eta\aleph}\geq b_{\eta\aleph}$, whil this vacuum background induces spontaneous Lorentz symmetry breaking through Kalb Ramond field self-interactions~\cite{altschul2010lorentz}. It is important to mention that when the potential depends on the scalar combination $\mathcal{B}_{\eta\aleph}\mathcal{B}^{\eta\aleph}$, the structure remains invariant with the scenario of locally Lorentz transformations, however gravitational action incorporating a non-minimally coupled Kalb Ramond field takes the following form~\cite{altschul2010lorentz,lessa2021traversable}.
\begin{widetext}
\begin{align}\label{KR1}&
\delta_{\text{Kalb Ramond}}^{\text {nonmin }}=\int \left(\frac{R}{2 \kappa}-\frac{1}{12} \mathcal{H}_{\lambda \eta \aleph} \mathcal{H}^{\lambda \eta \aleph}-V\left(\mathcal{B}_{\eta \aleph} \mathcal{B}^{\eta \aleph} \pm b_{\eta \aleph} b^{\eta \aleph}\right)+\frac{1}{2 \kappa}\left(\xi_2 \mathcal{B}^{\lambda \aleph} \mathcal{B}_\aleph^\eta R_{\lambda \eta}+\xi_3 \mathcal{B}^{\eta \aleph} \mathcal{B}_{\eta \aleph} R\right)+\mathcal{L}^M\right)e d^4 x,
\end{align}
\end{widetext}
In the above action, $e$ is associated with the metric determinant, while the constant $\kappa=8\pi G$ characterizes the strength of the gravitational interaction. The parameters $\xi_2$ and $\xi_3$ quantify the non-minimal interaction between the Kalb Ramond sector and spacetime geometry, and both carry the dimension of length squared, i.e., $[\xi]=L^2$. The action consists of three main parts: the curvature contribution arising from the Einstein-Hilbert sector, the dynamical contribution governed by the Kalb Ramond field strength, and the potential sector. The latter is particularly important, since it drives the field toward a non-zero vacuum configuration and thereby provides a mechanism for spontaneous Lorentz-symmetry breaking. The present study investigates imprints of a WH configuration in a regime where Lorentz symmetry is broken spontaneously. For this purpose, the Kalb Ramond field is taken in its vacuum state, specified through the condition involving the scalar combination $\mathcal{B}_{\eta\aleph}\mathcal{B}^{\eta\aleph}\pm b_{\eta\aleph}b^{\eta\aleph}$. In flat spacetime, one may introduce a constant Lorentz-violating vacuum expectation value $b_{\eta\aleph}$ satisfying $\partial_\mu b_{\eta\aleph}=0$. Under this assumption, the invariant quantity
\[
b^2=\Psi^{\eta\aleph}\Psi^{\varsigma\beta}b_{\eta\varsigma}b_{\aleph\beta}
\]
remains fixed, so that the Lorentz-violating background coefficients can be consistently described throughout spacetime by means of $b_{\eta\aleph}$. Furthermore, a constant background tensor $b_{\eta\aleph}$ gives rise to a vanishing Kalb Ramond field strength, namely $h_3=db_2=0$, and hence the corresponding vacuum Hamiltonian also vanishes~\cite{altschul2010lorentz}.

In curved spacetime, a direct generalization is obtained by imposing the covariant condition $\nabla_\mu b_{\eta\aleph}=0$, which likewise guarantees a zero contribution from the Kalb Ramond Hamiltonian~\cite{lessa2020modified}. Alternatively, the vacuum expectation value of the Kalb Ramond field may be characterized by requiring the background tensor to possess a constant norm, $b^2=b_{\eta\aleph}b^{\eta\aleph}$. This condition is associated with $b^{\eta\aleph}\nabla_\mu b_{\eta\aleph}=0$ and results in the disappearance of the potential term~\cite{casana2018exact}. In what follows, we adopt a Kalb Ramond vacuum configuration with constant norm. Let us now consider a pseudo electric field configuration, which can be conveniently expressed in the following form:
\begin{align}\label{KR2}&
b_2=-\widetilde{E}\left(x^1\right) d x^0 \wedge d x^1 .
\end{align}

Equation~\eqref{KR2} shows that the vacuum expectation value may be written as
$b_2=d\widetilde{\mathcal{A}}_1$, where
$\widetilde{A}_1=\widetilde{\mathcal{A}}_0(x^1)dx^0$ denotes the corresponding pseudo vector potential, and the associated pseudo electric component is defined by
$\widetilde{E}=-\partial_1\widetilde{\mathcal{A}}_0$. It is worth emphasizing that this vacuum configuration effectively represents a background electric type field while keeping the Hamiltonian contribution zero.

By varying the action given in Eq.~\eqref{KR1} with respect to the metric tensor, one obtains the modified Einstein field equations as~\cite{altschul2010lorentz,lessa2021traversable,casana2018exact}
\begin{align}\label{KR3}&
\mathcal{G}_{\eta \aleph}=R_{\eta \aleph}-\frac{1}{2} R g_{\eta \aleph}=\kappa T_{\eta \aleph}^{\xi_2}+\kappa T_{\eta \aleph}^{(M)}
\end{align}
where $T_{\eta\aleph}^{(M)}$ denotes the energy-momentum tensor (EMT), while
\begin{widetext}
\begin{align}\nonumber&
T_{\eta \aleph}^{\xi_2}=  \frac{\xi_2}{\kappa}\left(\frac{1}{2} g_{\eta \aleph} \mathcal{B}^{\varsigma \gamma} \mathcal{B}_\gamma^\beta R_{\varsigma \beta}-\mathcal{B}_\eta^\varsigma \mathcal{B}_\aleph^\beta R_{\varsigma \beta}-\mathcal{B}^{\varsigma \beta} \mathcal{B}_{\eta \beta} R_{\aleph \varsigma}-\mathcal{B}^{\varsigma \beta} \mathcal{B}_{\aleph \beta} R_{\eta \varsigma}+\frac{1}{2} \mathcal{D}_\varsigma \mathcal{D}_\eta\left(\mathcal{B}_{\aleph \beta} \mathcal{B}^{\varsigma \beta}\right)+\frac{1}{2} \mathcal{D}_\varsigma \mathcal{D}_\aleph\left(\mathcal{B}_{\eta \beta} \mathcal{B}^{\varsigma \beta}\right)\right. \\\label{KR4}
& \left.-\frac{1}{2} \mathcal{D}^2\left(\mathcal{B}_\eta^\varsigma \mathcal{B}_{\varsigma \aleph}\right)-\frac{1}{2} g_{\eta \aleph} \mathcal{D}_\varsigma \mathcal{D}_\beta\left(\mathcal{B}^{\varsigma \gamma} \mathcal{B}_\gamma^\beta\right)\right)
\end{align}
\end{widetext}
We next formulate the field equations for a WH spacetime sourced by the EMT contribution arising from the Lorentz-violating Kalb Ramond vacuum configuration given in Eq. \eqref{KR4}.

\section{FIELD EQUATIONS}

To construct WH configurations in the Kalb Ramond gravitational framework, now we proceed by considering the Morris-Thorne spacetime describing a static and spherically symmetric traversable WH configuration.
\begin{align}\label{KR5}
ds^2=-e^{2\Phi(r)}dt^2+\left(1-\frac{Y(r)}{r}\right)^{-1}dr^2+r^2d\theta^2+r^2\sin^2\theta\,d\phi^2 .
\end{align}
Here, $\Phi(r)$ denotes the redshift function, whereas $Y(r)$ is shape function of geometry. For a physically admissible traversable WH, the redshift $\Phi(r)$ must remain finite everywhere outside and at the throat in order to avoid the formation of horizons. The throat is located at $r=r_0$, where the shape function satisfies $Y(r_0)=r_0$. In addition, the shape function is required to obey the flare-out condition $Y'(r)<1$ at the throat, along with the condition $Y(r)/r<1$ for $r>r_0$. Throughout this work, a prime denotes differentiation with respect to the radial coordinate $r$.

Using the Kalb Ramond vacuum expectation value ansatz introduced in Eq. \eqref{KR2}, together with the constant norm condition $b^2=g^{\eta\varsigma}g^{\aleph\beta}b_{\eta\aleph}b_{\varsigma\beta},$ one obtains the following form of the field component $\widetilde{E}(r)$ for the spacetime geometry described by Eq.~\eqref{KR5}:
\begin{align}\label{KR6}
\tilde{E}(r)=\frac{|b| e^{\phi(r)}}{\sqrt{2\left(1-\frac{Y(r)}{r}\right)}},
\end{align}
here $b$ stand as a constant parameter, however it is noteworthy that the radial pseudo electric background configuration
\[
\widetilde{E}^{\eta}=(0,\widetilde{E},0,0)
\]
is compatible with both the static character and spherical symmetry of the spacetime which follows from the orthogonality of the background vector $\widetilde{E}^{\eta}$ to the timelike Killing vector
$t^\eta=\left(\frac{\partial}{\partial t}\right)^\eta$
as well as to the azimuthal spacelike Killing vector
$\psi^\eta=\left(\frac{\partial}{\partial \phi}\right)^\eta$.
For the WH line element \eqref{KR5} above expressions \eqref{KR3} take the following form:
\begin{align}\label{KR7}
\mathcal{G}_{t t} & =\frac{\lambda}{4}\left[3 R_{t t}-\left(1-\frac{Y(r)}{r}\right) R_{r r}\right]+\kappa T_{t t}^M, \\\label{KR8}
\mathcal{G}_{r r} & =\frac{\lambda}{4}\left[3 R_{r r}-\left(1-\frac{Y(r)}{r}\right)^{-1} R_{t t}\right]+\kappa T_{r r}^M,  \\\label{KR9}
\mathcal{G}_{\theta \theta} & =\frac{\lambda r^2}{4}\left[R_{t t}-\left(1-\frac{Y(r)}{r}\right) R_{r r}\right]+\kappa T_{\theta \theta}^M,  \\\label{KR10}
\mathcal{G}_{\phi \phi} & =\sin ^2 \theta G_{\theta \theta},
\end{align}
in above expression $\lambda=|b|^2\xi_2$ denotes the Kalb Ramond gravitational parameter. Here, we restrict our attention to the zero tidal force case by taking the redshift function to be constant, namely $\Phi(r)=A=\text{constant}$, while within the scenario of this assumption, the corresponding non-vanishing components of Ricci tensor $R_{\eta\aleph}$ take the following explicit form:
\begin{align}\label{KR11}&
R_{r r}=\frac{Y^{\prime}(r)}{1-\frac{Y(r)}{r}}-\frac{Y(r)}{r\left(1-\frac{Y(r)}{r}\right)}, \quad R_{\theta \theta}=\frac{Y(r)}{2 r}+\frac{Y^{\prime}(r)}{2}.
\end{align}
In the background of the Kalb Ramond field an anisotropic EMT is given the following for;
\begin{align}\label{KR12}&
\left(T_\aleph^\eta\right)^{(M)}=\left(\rho+P_{rad}\right) u^\eta u_\aleph+P_{rad} g_\eta^\aleph+\left(P_{tan}-P_{rad}\right) \chi^\eta \chi_\aleph.
\end{align}
Here, $\rho$ represents the energy density, while $P_{rad}$ and $P_{tan}$ denote the radial and tangential pressures, respectively. Moreover, $u^\eta$ and $\chi^\eta$ characterize the timelike and spacelike directions of the anisotropic fluid.
Accordingly, the mixed components of the matter EMT $\left(T^\eta_{\aleph}\right)^{(M)}$ can be written as
\begin{align}\label{KR13}&
\left(T_t^t\right)^{(M)}=-\rho, \quad\left(T_r^r\right)^{(M)}=P_{rad}, \quad\left(T_\theta^\theta\right)^{(M)}=\left(T_\phi^\phi\right)^{(M)}=P_{tan}.
\end{align}
Therefore, using gravitational units Eqs.~(7)-(10) may be expressed as
\begin{align}\label{KR14}
\rho &=\frac{1}{8 \pi}\left[\frac{Y^{\prime}(r)}{r^2}+\frac{\lambda}{4 r^3}\left(r Y^{\prime}(r)-Y\right)\right],  \\\label{KR15}
P_{rad} &=-\frac{1}{8 \pi}\left[\frac{Y(r)}{r^3}+\frac{3 \lambda}{4 r^3}\left(r Y^{\prime}(r)-Y(r)\right)\right],  \\\label{KR16}
P_{tan} & =-\frac{1}{8 \pi}\left(1-\frac{\lambda}{2}\right) \frac{r Y^{\prime}(r)-Y(r)}{2 r^3} .
\end{align}

It is noticed that in the above equations, there are four unknowns $\rho$,~$P_ {rad}$,~$P_ {tan}$, and $Y(r)$. To discover the solution to the WH problem, various methods can be employed due to the surplus of unknowns compared to the number of equations. Below, we present the proposed action plan to address the field equations. It is very interesting to analyze at which spacetime arena, the energy conditions are breached within time-independent, spherically symmetric, $4D$ spacetime configurations. This violation of energy conditions is a notable phenomenon within this specific spacetime framework. The flaring out condition determines whether these conditions are violated \cite{morris1988wormholes}. Furthermore, in higher dimensional theories, the only places where energy conditions violations can be avoided or satisfied may be near the throat of the WH.

\section{Determination of the WH Shape Function from the Einasto Profile and Basic Criteria Checks}\label{ShapeFunction}

A thorough understanding of large scale cosmic structures, such as dark matter halos surrounding galaxy clusters, requires accurate modeling of their density distributions, while cosmological N-body simulations suggest that dark matter halos can be effectively represented using three parameter density profiles \cite{merritt2006empirical,hayashi2008understanding}. Among these, the Einasto-density profile emerged as a powerful tool for describing dark matter halos \cite{gao2008redshift,navarro2010diversity,de2019estimation}. The Einasto-density profile also found applications in the study of electromagnetic fields and fuzzy WHs \cite{almatroud2025electromagnetic}, as well as in the construction of WHs using various dark matter density distributions \cite{yousaf2025wormholes}. Furthermore, it provides insights into the central regions of spiral galaxies and the formation of early cosmic structures, including anisotropic WHs analyzed via the Minimal Geometric Deformation technique in the presence of dark matter halos \cite{gadotti2009structural,almatroud2025decoupling}, whereas the three-parameter, non-singular Einasto-density profile offers an accurate and flexible representation of dark matter halos \cite{navarro1997universal}. The literature offers several established methodologies for developing and examining WH models \cite{hawking1988wormholes,singh2020conformally,hassan2021traversable,yousaf2023cylindrical}. The Einasto-density profile is defined through its logarithmic slope, which follows a power law form \cite{retana2012analytical,baes2022einasto}:
$\mathcal{X}(r) = d \ln \rho/d \ln r \propto r^{{1/n}},$
where $n$ is the Einasto index, a free parameter controlling the curvature of the profile. Integrating this slope leads to the general density function:
$\ln \left(\rho(r)/\rho_s \right) = -a_n \left( \left( r/r_s \right)^{1/n} - 1 \right),$
where $r_s$ denotes the scale radius containing half of the total mass, $a_n$ is a dimensionless parameter ensuring proper normalization, $\rho_s$ is the density at $r_s$, and the central density is $\rho_s = \rho_s e^{a_n}$. A commonly used representation of the Einasto-density profile for dark matter halos is:
$\ln \left(\rho(r)/\rho_{-2} \right) = -2 \gamma_1 \left( \left( r/r_{-2} \right)^{1/n} - 1 \right),$
where $r_{-2}$ and $\rho_{-2}$ correspond to the radius and density at which the logarithmic slope of the density equals $-2$, so equivalently, Einasto profile defined as
\begin{align}\label{a15}
\rho(r) = \rho_s e^{-\left(\frac{r}{r_s}\right)^{\frac{1}{\mathit{n}}}},
\end{align}
For the Einasto-density profile to accurately represent real galactic structures, the density function and associated quantities must satisfy certain physical conditions \cite{einasto1969galactic}: $\rho(r)$ must be strictly positive and finite throughout the spatial domain, while the density should smoothly decrease to zero as $r \to \infty$, avoiding unphysical asymptotic behavior and the total mass, effective radius, and multipole expansions must be finite, whereas all descriptive functions, including mass profiles and gravitational potential, should be continuous without sudden jumps or discontinuities. In the context of traversable WHs, the presence of exotic matter leads to a violation of the null energy constraints. The stability of such WHs can be analyzed either by gradual collapse or by considering configurations with sufficiently large throats, however to derive the corresponding shape function $Y(r)$, we combine the Einstein field equation with the Einasto-density profile from Eq.~\eqref{a15}, leading to the differential equation:
\begin{align}\label{a18}
\frac{1}{8 \pi}\left[\frac{Y^{\prime}(r)}{r^2}+\frac{\lambda}{4 r^3}\left(r Y^{\prime}(r)-Y(r)\right)\right] =\rho_s e^{-\left(\frac{r}{r_s}\right)^{\frac{1}{\mathit{n}}}}.
\end{align}
Solving Eq.~\eqref{a18} yields the WH shape function:
\begin{widetext}
\begin{align}\label{a19}
Y(r) = c r^{\frac{\lambda }{\lambda +4}}-\frac{32 \pi  \mathit{n} \rho_s r^{\frac{\lambda }{\lambda +4}+\frac{2 (\lambda +6)}{\lambda +4}} \left(\left(\frac{r}{r_s}\right)^{\frac{1}{\mathit{n}}}\right)^{-\frac{2 (\lambda +6) \mathit{n}}{\lambda +4}} \Gamma \left(\frac{2 \mathit{n} (\lambda +6)}{\lambda +4},\left(\frac{r}{r_s}\right)^{\frac{1}{\mathit{n}}}\right)}{\lambda +4},
\end{align}
\end{widetext}
where $c$ is the integration constant. Applying the boundary condition at WH throat, $Y(r_0) = r_0$, determines $c$ as
\begin{align}\label{a20}
c = r_0^{\frac{4}{\lambda +4}} \left(\frac{32 \pi  \mathit{n} \rho_s r_0^2 \left(\left(\frac{r_0}{r_s}\right){}^{\frac{1}{\mathit{n}}}\right){}^{-\frac{2 (\lambda +6) \mathit{n}}{\lambda +4}} \Gamma \left(\frac{2 \mathit{n} (\lambda +6)}{\lambda +4},\left(\frac{r_0}{r_s}\right){}^{\frac{1}{\mathit{n}}}\right)}{\lambda +4}+1\right).
\end{align}
This derivation provides a self consistent expression for the WH shape function $Y(r)$ based on the Einasto density profile, ensuring a physically meaningful and mathematically robust WH geometry suitable for further analysis of stability and energy conditions.
The ultimate form of $Y(r)$ (i.e. shape function) is provided as,
\begin{widetext}
\begin{align}\nonumber&
Y(r)=\frac{1}{\lambda +4}\left(r^{\lambda /(\lambda +4)} \left(32 \pi  \mathit{n} \rho_s \left(r_0^{2 (\lambda +6)/(\lambda +4)} \left(\left(r_0/r_s\right){}^{\frac{1}{\mathit{n}}}\right){}^{-2 (\lambda +6) \mathit{n}/(\lambda +4)} \Gamma \left(2 \mathit{n} (\lambda +6)/(\lambda +4),\left(r_0/r_s\right){}^{\frac{1}{\mathit{n}}}\right) \right.\right.\right.
\\\label{shapefunction}& \left.\left.\left. -r^{2 (\lambda +6)/(\lambda +4)} \left(\left(r/r_s\right)^{\frac{1}{\mathit{n}}}\right)^{-2 (\lambda +6) \mathit{n}/(\lambda +4)}\Gamma \left(2 \mathit{n} (\lambda +6)/(\lambda +4),\left(r/r_s\right)^{\frac{1}{\mathit{n}}}\right)\right)+(\lambda +4) r_0^{4/(\lambda +4)}\right)\right).
\end{align}
\end{widetext}
\begin{figure*}[htbp]
\centering
{{\includegraphics[height=2.0 in, width=3.2 in]{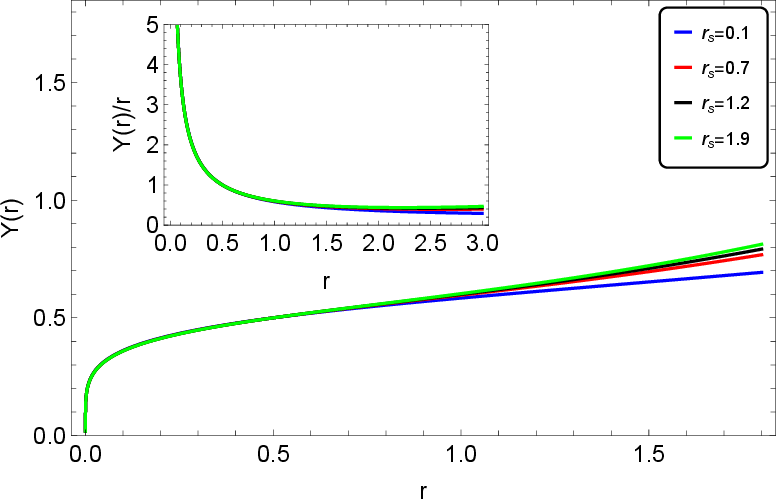}}}
\qquad
{{\includegraphics[height=2.0 in, width=3.2 in]{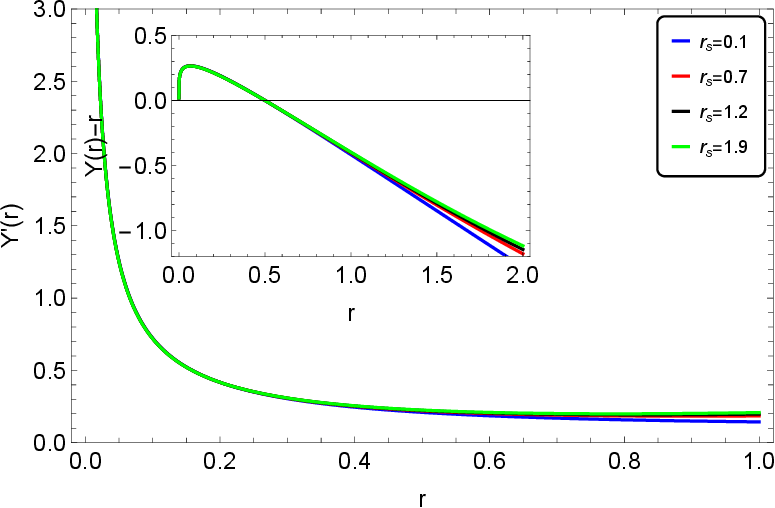}}}
\qquad
{{\includegraphics[height=2.0 in, width=3.2 in]{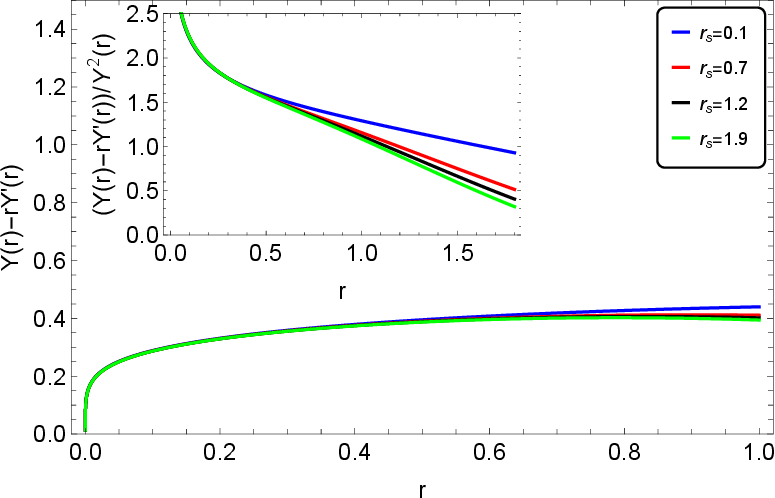}}}
\caption{Plots of shape function \eqref{shapefunction} at 0.6, variation of the shape function itself plotted in main panel (a), inner small panel (a) illustrates the normalized ratio of \eqref{shapefunction} to radial coordinate. Main panel (b) presents the derivative of \eqref{shapefunction} indicating its rate of change, and difference between the shape function and the radial coordinate highlighting the deviation from a linear profile is in inner small panel (b). While difference between the shape function and the product of the radial coordinate with its derivative capturing the interplay of linear and nonlinear effects in main panel (c) and inner small panel (c) shows the normalized flare-out condition.}
\label{flareout1}
\end{figure*}

The main and inset panels of Figs.~\ref{flareout1} provide a detailed graphical analysis of the
shape function constructed from the Einasto dark matter density profile as defined in Eq.~\eqref{a15} within
the Kalb Ramond gravitational scenario. Panel (a) of Figs.~\ref{flareout1} displays the variation of
$Y(r)$ with respect to the radial coordinate $r$ for different values of the
Einasto scale radius $r_s$, however, the shape function remains positive as well as varies
smoothly throughout the considered radial domain, indicating a regular
WH geometry. The inset of panel (a) of Figs.~\ref{flareout1} depicts the ratio $Y(r)/r$, which
decreases with increasing radial distance and satisfies the fundamental
condition
\[
\frac{Y(r)}{r}<1, \qquad r>r_0,
\]
outside the throat.
Panel (b) of Figs.~\ref{flareout1} illustrates the first derivative $Y'(r)$ for different values of
$r_s$, which provides a direct test of the flare-out condition at the throat, whereas for the adopted values of the Einasto and Kalb Ramond model parameters, the
condition $Y'(r_0)<1$ is satisfied, confirming the geometrical viability of
the constructed WH configuration. The inset of panel (b) shows the
quantity $Y(r)-r$, which vanishes at the throat and becomes negative for
$r>r_0$, consistently demonstrating that $Y(r)<r$ outside the throat.
Furthermore, panel (c) of Figs.~\ref{flareout1} presents the combination $Y(r)-rY'(r)$, while its
inset shows the normalized quantity
\[
\frac{Y(r)-rY'(r)}{Y^2(r)}.
\]
The positive behavior of these quantities in the relevant radial region
provides an additional graphical confirmation of the flare-out property of
the WH geometry. The dependence of the curves on $r_s$ demonstrates
that the Einasto scale radius modifies the detailed radial behavior of the
shape function, whereas the Kalb Ramond gravitational contribution is
incorporated through the parameter $\lambda$ appearing explicitly in
$Y(r)$. Consequently, the graphical results confirm that the obtained
Einasto supported WH solutions satisfy the essential geometrical
requirements for traversability within the considered Kalb Ramond gravity scenario.

\section{Geometrical Analysis of WH Embedding Structure}
\label{EmbeddingDiagrams}

\begin{figure*}[htbp]
\centering
{{\includegraphics[height=2.6 in, width=2.8 in]{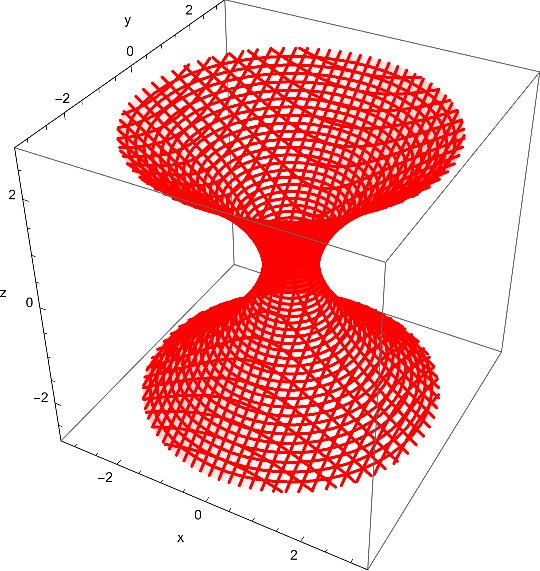}}}
\qquad
{{\includegraphics[height=2.6 in, width=2.8 in]{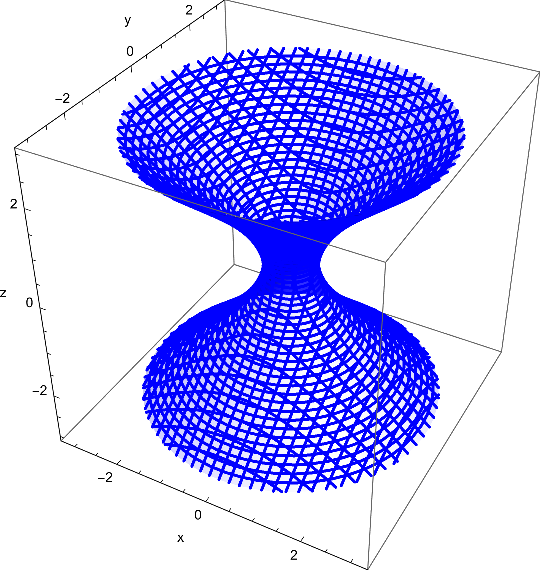}}}
\qquad
{{\includegraphics[height=2.6 in, width=2.8 in]{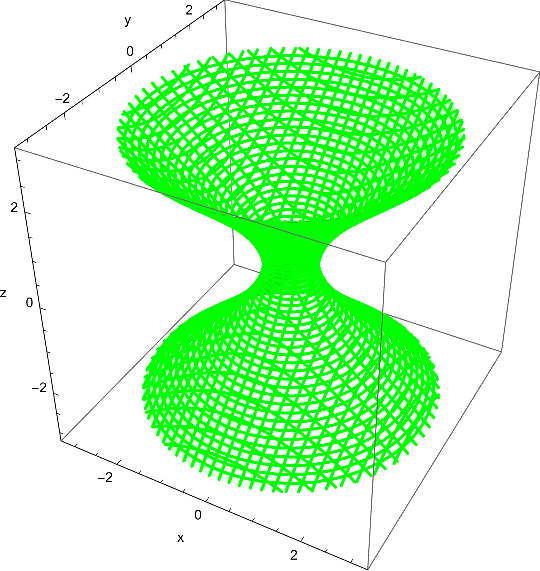}}}
\qquad
{{\includegraphics[height=2.6 in, width=2.8 in]{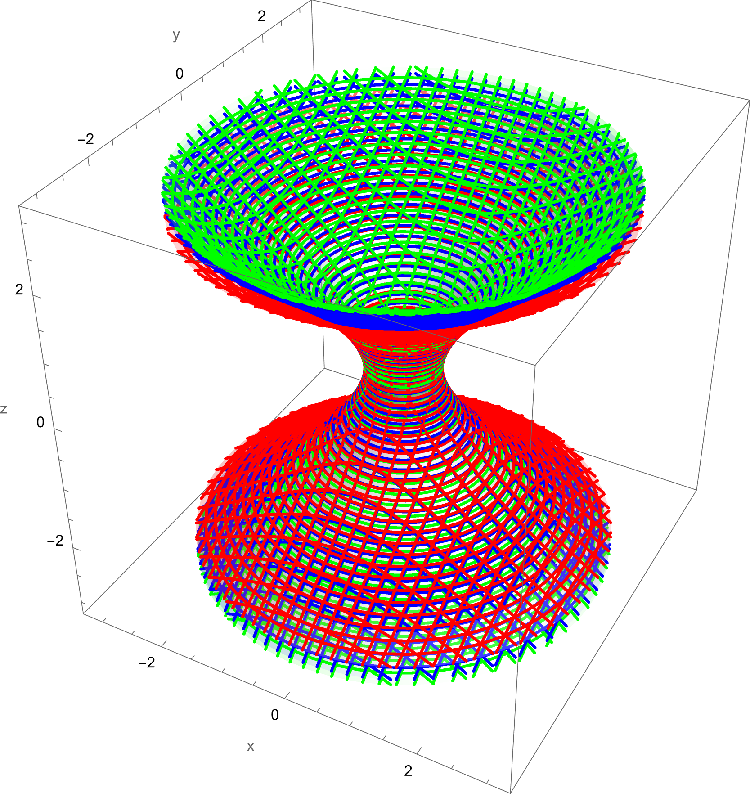}}}
\caption{3D embedding diagrams of the shape function \eqref{shapefunction} with throat radius \( r_0=0.5 \).}
\label{fig:combined3D}
\end{figure*}

To obtain a clear geometrical interpretation of the constructed WH configuration, we analyze the corresponding embedding surfaces for representative choices of the parameters associated with the Einasto matter distribution and the Kalb Ramond gravitational sector, since the shape function $Y(r)$ obtained in the present model contains the incomplete Gamma function, the embedding geometry provides a particularly useful visualization of the spatial curvature, flare out behavior, as well as throat structure of the resulting WH spacetime. In the present scenario, the geometry is characterized by the Einasto parameters $\rho_s$, $r_s$, and $n$, the throat radius $r_0$, and the Kalb Ramond gravitational parameter $\lambda$ entering the shape function. Owing to the static and spherically symmetric nature of the spacetime, the intrinsic spatial geometry can be conveniently examined on a constant time hypersurface, $t=\mathrm{constant}$, furthermore, restricting the analysis to the equatorial plane, $\theta=\pi/2$, reduces the spatial geometry to an effectively two dimensional surface without losing the essential geometrical information associated with the WH throat. Under these assumptions, the spatial line element becomes
\begin{equation}
ds^2=
\left(1-\frac{Y(r)}{r}\right)^{-1}dr^{2}
+r^2 d\phi^2,
\label{eq:embedmetric}
\end{equation}
where the throat is located at $r=r_0$ and satisfies the fundamental condition
$Y(r_0)=r_0$. Here, $r$ denotes the circumferential radial coordinate and $\phi$ is the azimuthal coordinate. To visualize the intrinsic curvature described by Eq.~\eqref{eq:embedmetric}, the two dimensional spatial section is embedded in a three dimensional Euclidean space with Cartesian coordinates $(x,y,z)$, whereas the corresponding parametrization may be written as
$x=r\cos\phi, ~~y=r\sin\phi, ~~z=z(r).$
In this representation, the function $z(r)$ determines the embedding profile and directly characterizes the manner in which the WH surface flares outward from its minimum radius at the throat, while the Euclidean line element in cylindrical coordinates takes the form
\begin{equation}
\label{34}
ds^2=dr^2+r^2d\phi^2+dz^2.
\end{equation}
For an embedded surface described by $z=z(r)$, one has
$dz=\frac{dz}{dr}\,dr,$ and hence the induced metric on the embedded surface becomes
\begin{equation}
ds^2=
\left[
1+\left(\frac{dz}{dr}\right)^2
\right]dr^2+r^2d\phi^2.
\end{equation}
Comparing the above induced metric with Eq.~\eqref{eq:embedmetric}, we obtain the differential equation governing the embedding profile,
\begin{equation}
\frac{dz}{dr}
=
\pm
\left(
\frac{r}{Y(r)}-1
\right)^{-1/2},
\qquad
z(r_0+\epsilon)=0,
\label{eq:embeddingDE}
\end{equation}
where $r_0$ denotes the WH throat radius and $\epsilon$ is a sufficiently small positive quantity introduced to initiate the numerical integration slightly outside the coordinate singularity at the throat, however the positive and negative branches of Eq.~\eqref{eq:embeddingDE} generate the upper and lower embedding sheets, respectively, which join at the throat and represent the two spatial regions connected by the WH geometry. An important geometrical feature of Eq.~\eqref{eq:embeddingDE} appears at the throat, since $Y(r_0)=r_0$, the denominator of the embedding equation vanishes at $r=r_0$, leading to
\begin{equation}
\left|\frac{dz}{dr}\right|\rightarrow\infty
\quad
\text{as}
\qquad
r\rightarrow r_0^{+}.
\end{equation}
Thus, the embedding curve develops a vertical tangent at the throat, which is the characteristic geometrical manifestation of the WH flare out structure, while moving away from the throat, the magnitude of $dz/dr$ decreases progressively. For an asymptotically flat configuration satisfying
\begin{equation}
\frac{Y(r)}{r}\rightarrow 0
\quad
\text{as}
\quad
r\rightarrow\infty,
\end{equation}
the embedding slope approaches zero,
$\frac{dz}{dr}\rightarrow 0,$
indicating that the embedded surface gradually approaches a flat geometry at sufficiently large radial distances.
The embedding function can therefore be expressed formally as
\begin{equation}
\label{37}
z(r)
=
\pm
\int_{r_0}^{r}
\left[
\frac{\bar r}{Y(\bar r)}-1
\right]^{-1/2}
d\bar r,
\end{equation}
where the auxiliary integration variable $\bar r$ introduced to distinguish it from the upper integration limit $r$, because the explicit form of $Y(r)$ contains incomplete Gamma functions, the integral in Eq.~\eqref{37} is evaluated numerically for the selected model parameters. The resulting embedding diagrams are displayed in Fig.~\ref{fig:combined3D}. These graphical representations provide a direct visualization of the two-sheeted WH geometry and illustrate how the spatial curvature and radial extension of the throat region respond to variations in the parameters governing the Einasto matter distribution and the Kalb Ramond gravitational contribution. The embedding diagrams corresponding to the obtained WH geometry are
presented in Figs.~2 and 3 which provide a direct geometrical
visualization of the influence of the Einasto matter distribution and the
Kalb Ramond gravitational contribution on the spatial structure of the
WH. For the adopted parameter choices, the upper and lower embedding
branches join smoothly at the throat, thereby exhibiting the characteristic
two sheeted structure of a traversable WH. The variation of the model
parameters modifies the radial extension as well as curvature of the embedded
surface without destroying the throat structure. In particular, changes in
the Einasto parameters alter the matter distribution entering the shape
function, whereas the Kalb Ramond parameter $\lambda$ modifies the
gravitational contribution to the resulting geometry. Nevertheless, the
constructed configurations preserve the essential flare-out behavior in
the considered parameter domain.

\begin{figure*}[htbp]
\centering
{{\includegraphics[height=2.8 in, width=4.0 in]{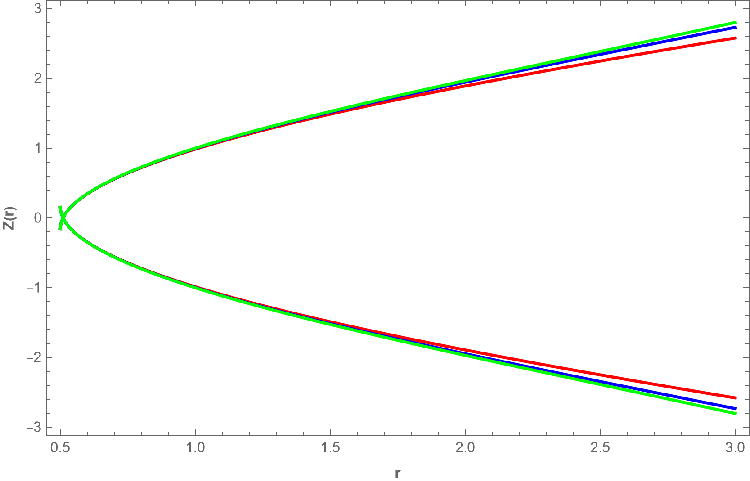}}}
\caption{2D embedding diagram for parameter set and throat radius \( r_0=0.5 \).}
\label{fig:combined2D}
\end{figure*}
The three dimensional embedding diagrams show that the WH geometry
remains regular for the considered choices of the model parameters, while the
embedded surfaces exhibit a pronounced minimum radius at the throat as well as
flare outward on both sides, providing a direct geometrical confirmation of
the traversable WH structure. Differences among the plotted
configurations reflect the sensitivity of the shape function to the Einasto
matter parameters and to the Kalb Ramond gravitational parameter
$\lambda$, however these parameters control the curvature as well as radial extension of
the embedding surface, while the qualitative two-sheeted WH geometry
is preserved. Moreover, the gradual flattening of the embedding profiles
away from the throat is consistent with the weakening of the geometrical
deformation at larger radial distances.
In this section, we construct and analyze the embedding surfaces associated
with the obtained shape function, while the embedding geometry provides a useful
geometrical representation of the WH throat and allows us to examine
how the spatial curvature responds to the characteristic parameters of the
Einasto matter distribution and the Kalb Ramond gravitational sector.
Accordingly, the geometry is governed by the parameters
$\rho_s$, $r_s$, $n$, $r_0$, and the Kalb Ramond parameter $\lambda$
appearing in the shape function.
The WH metric takes the following form for the previously derived shape function.
\begin{widetext}
\begin{align}\nonumber&
ds^2=-dt^2+\left(1-\frac{1}{r(\lambda +4)}\left(r^{\frac{\lambda }{\lambda +4}} \left(32 \pi  \mathit{n} \rho_s \left(r_0^{\frac{2 (\lambda +6)}{\lambda +4}} \left(\left(\frac{r_0}{r_s}\right){}^{\frac{1}{\mathit{n}}}\right){}^{-\frac{2 (\lambda +6) \mathit{n}}{\lambda +4}} \Gamma \left(\frac{2 \mathit{n} (\lambda +6)}{\lambda +4},\left(\frac{r_0}{r_s}\right){}^{\frac{1}{\mathit{n}}}\right)-r^{\frac{2 (\lambda +6)}{\lambda +4}}  \right.\right.\right.\right.
\\\label{metric2}& \left.\left.\left.\left.\left. \times\left(\left(\frac{r}{r_s}\right)^{\frac{1}{\mathit{n}}}\right)^{-\frac{2 (\lambda +6) \mathit{n}}{\lambda +4}}\Gamma \left(\frac{2 \mathit{n} (\lambda +6)}{\lambda +4},\left(\frac{r}{r_s}\right)^{\frac{1}{\mathit{n}}}\right)\right)+(\lambda +4) r_0^{\frac{4}{\lambda +4}}\right)\right)\right)\right)^{-1}dr^2+r^2d\theta^2+r^2\sin^2\theta\,d\phi^2 .
\end{align}
\end{widetext}

\section{Some Physical Features and Energy Constraints of Fuzzy WHs}\label{PhysicalFeatures_EnergyConstrains}

\begin{figure*}[htbp]
\centering
{{\includegraphics[height=2.0 in, width=3.2 in]{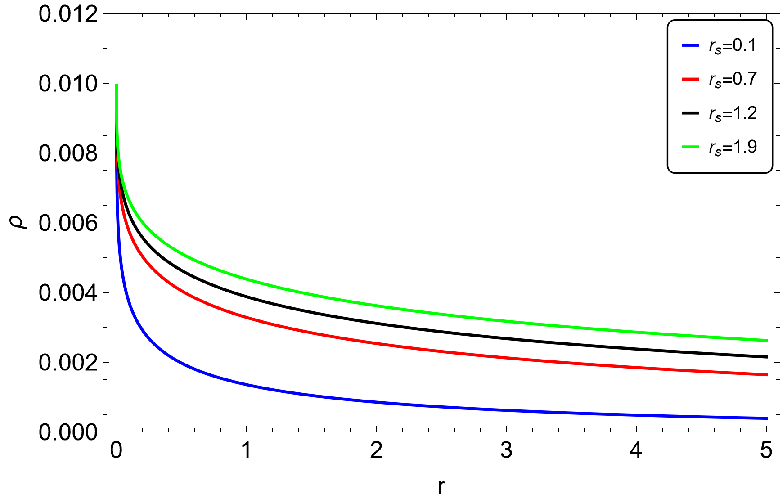}}}
\qquad
{{\includegraphics[height=2.0 in, width=3.2 in]{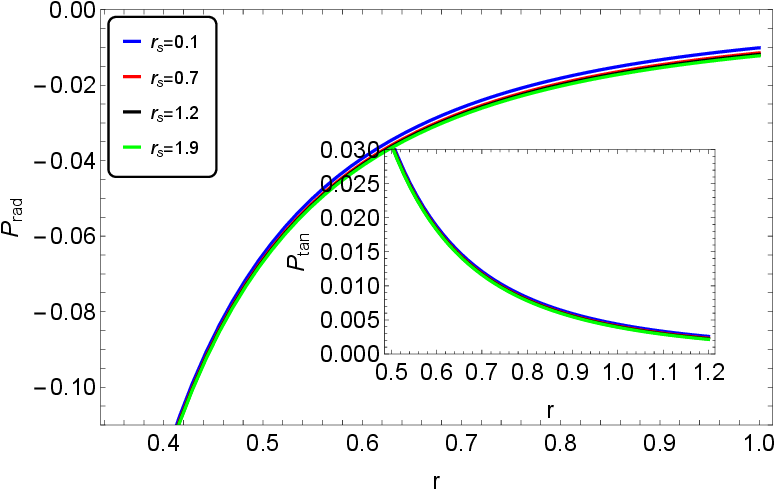}}}
\caption{Behavior of $\rho(r)$ (panel (a)), $P_{rad}(r)$, and $P_{tan}(r)$ (panel (b)) at $r_0=0.5$, $\rho_s $=0.3(Line), $\rho_s $=0.5(Dashing), $\rho_s $=0.7(Dotted), $\rho_s $=0.9(Dashed), and $n=3.33$.}
\label{energydensity}
\end{figure*}

\begin{figure*}[htbp]
\centering
{{\includegraphics[height=2.0 in, width=3.2 in]{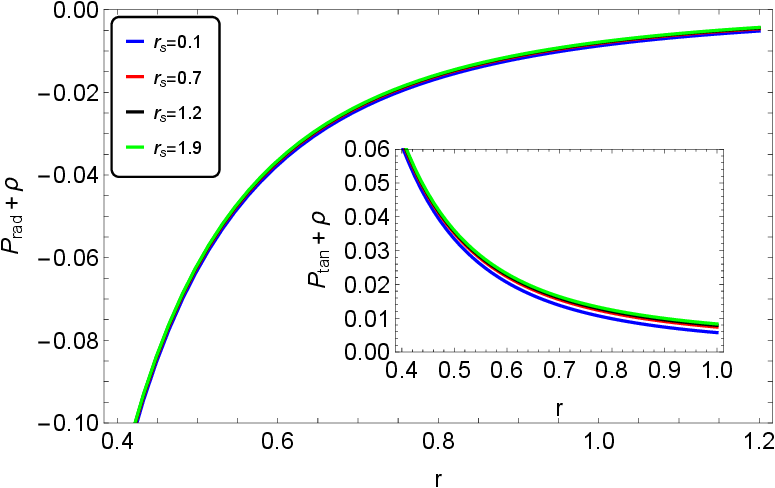}}}
\qquad
{{\includegraphics[height=2.0 in, width=3.2 in]{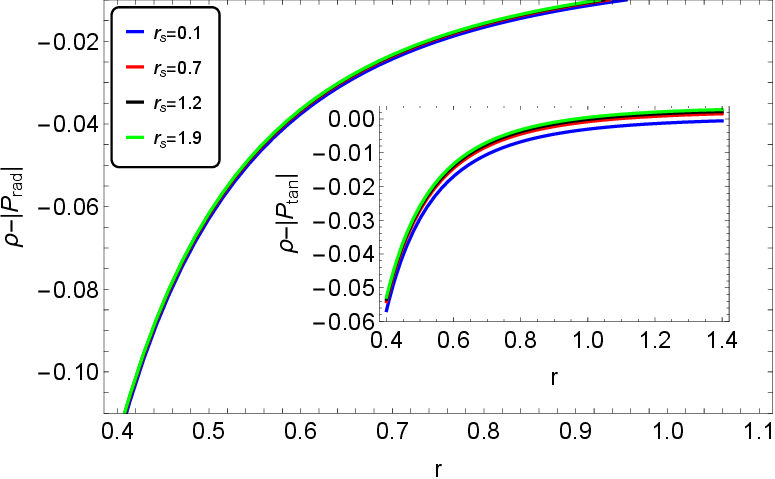}}}
\caption{Behavior of NECs (panel (a)) and DECs (panel (b)) at $r_0=0.5$, $\rho_s $=0.3(Line), $\rho_s $=0.5(Dashing), $\rho_s $=0.7(Dotted), $\rho_s $=0.9(Dashed), and $n=3.33$.}
\label{NEC}
\end{figure*}

Energy conditions play a fundamental role in determining the physical feasibility and matter distribution of WH geometries, whereas they provide mathematical constraints on how energy density and pressure behave in curved spacetime. For traversable WHs, the violation of the null energy constraint near the throat is typically unavoidable, as such violation prevents gravitational collapse and allows the WH to remain open, and therefore, examining the behavior of energy conditions helps identify the regions where exotic matter is required to sustain the WH structure.
In Einstein's general theory of relativity, energy conditions impose constraints on the EMT of the matter fields \cite{Yuennan2025Wormholes,Ditta2025MassiveGravity,yousaf2025interpretation} and the main energy conditions for an anisotropic fluid with energy density $\rho$, radial pressure $P_{rad}$, and tangential pressure $P_{tan}$ are summarized as follows.
\begin{itemize}
    \item Weak Energy Condition (WEC):
$T_{\eta\aleph}v^\eta v^\aleph \ge 0,$
where $v^\eta$ is any timelike vector. For anisotropic matter:
$\rho \ge 0,~\rho + P_{rad} \ge 0,~\rho + P_{tan} \ge 0.$
    \item Strong Energy Condition (SEC):
$\left(T_{\eta\aleph}-\frac{T}{2}g_{\eta\aleph}\right)v^\eta v^\aleph \ge 0,$
which yields
$\rho + P_{rad} + 2P_{tan} \ge 0,~\rho + P_{rad} \ge 0.$
    \item Null Energy Condition (NEC):
$T_{\eta\aleph}k^\eta k^\aleph \ge 0,$
for any null vector $k^\eta$, equivalent to:
$\rho + P_{rad} \ge 0,~\rho + P_{tan} \ge 0. $
    \item Dominant Energy Condition (DEC):
$T_{\eta\aleph}v^\eta v^\aleph \ge 0, \label{b12}$
with the anisotropic form:
$\rho \ge 0,~\rho \ge |P_{rad}|,~\rho \ge |P_{tan}|.$
\end{itemize}
These conditions impose strict physical constraints; however, in Einstein's general theory of relativity, constructing a traversable WH typically requires exotic matter that violates the NEC near the throat, whereas for the fuzzy WH configuration studied here, the matter variables $\rho$, $P_{rad}$, and $P_{tan}$ corresponding to the Einasto-density profile together with the constructed shape function are obtained as:
\begin{widetext}
\begin{align}\label{a23rho}
\rho &=\rho_s e^{-\left(\frac{r}{r_s}\right)^{\frac{1}{\mathit{n}}}},
\\\nonumber
P_ {rad}&=\frac{-1}{32 \pi  (\lambda +4)^2 r^3}\left(32 \pi  \rho_s \left(r^3 \left(3 \lambda  (\lambda +4) e^{-\left(\frac{r}{r_s}\right)^{\frac{1}{\mathit{n}}}}+8 (\lambda -2) \mathit{n} \left(\left(\frac{r}{r_s}\right)^{\frac{1}{\mathit{n}}}\right)^{-\frac{2 (\lambda +6) \mathit{n}}{\lambda +4}} \Gamma \left(\frac{2 \mathit{n} (\lambda +6)}{\lambda +4},\left(\frac{r}{r_s}\right)^{\frac{1}{\mathit{n}}}\right)\right)\right.\right.
\\\label{a23}&\left.\left.-8 (\lambda -2) \mathit{n} r^{\frac{\lambda }{\lambda +4}} r_0^{\frac{2 (\lambda +6)}{\lambda +4}} \left(\left(\frac{r_0}{r_s}\right){}^{\frac{1}{\mathit{n}}}\right){}^{-\frac{2 (\lambda +6) \mathit{n}}{\lambda +4}} \Gamma \left(\frac{2 \mathit{n} (\lambda +6)}{\lambda +4},\left(\frac{r_0}{r_s}\right){}^{\frac{1}{\mathit{n}}}\right)\right)-8 \left(\lambda ^2+2 \lambda -8\right) r^{\frac{\lambda }{\lambda +4}} r_0^{\frac{4}{\lambda +4}}\right),
\\\nonumber
P_{tan}&=\frac{-1}{16 \pi  (\lambda +4)^2 r^3}\left(\left(1-\frac{\lambda }{2}\right) \left(32 \pi  \rho_s \left(r^3 \left((\lambda +4) e^{-\left(\frac{r}{r_s}\right)^{\frac{1}{\mathit{n}}}}+4 \mathit{n} \left(\left(\frac{r}{r_s}\right)^{\frac{1}{\mathit{n}}}\right)^{-\frac{2 (\lambda +6) \mathit{n}}{\lambda +4}} \Gamma \left(\frac{2 \mathit{n} (\lambda +6)}{\lambda +4},\left(\frac{r}{r_s}\right)^{\frac{1}{\mathit{n}}}\right)\right)\right.\right.\right.
\\\label{a24}&\left.\left.\left.-4 \mathit{n} r^{\frac{\lambda }{\lambda +4}} r_0^{\frac{2 (\lambda +6)}{\lambda +4}} \left(\left(\frac{r_0}{r_s}\right){}^{\frac{1}{\mathit{n}}}\right){}^{-\frac{2 (\lambda +6) \mathit{n}}{\lambda +4}} \Gamma \left(\frac{2 \mathit{n} (\lambda +6)}{\lambda +4},\left(\frac{r_0}{r_s}\right){}^{\frac{1}{\mathit{n}}}\right)\right)-4 (\lambda +4) r^{\frac{\lambda }{\lambda +4}} r_0^{\frac{4}{\lambda +4}}\right)\right).
\end{align}
\end{widetext}
To examine the fuzzy WH structure, we focus on the regime $\xi \neq 1$ and Einasto index $n=3.33$, which controls the shape of the Einasto-density profile model, while the graphical analysis in Fig.~\ref{energydensity} shows the behavior of the density and pressures for $r_0=0.5$.
Panel (a) of Fig.~\ref{energydensity} shows that $\rho$ remains positive near the throat and decreases monotonically toward zero, indicating that WH supporting matter is concentrated around $r_0$. In panel (b), the radial pressure $P_{rad}$ is negative near the throat, consistent with the exotic matter requirement, whereas $P_{tan}$ is positive in this region. This anisotropic behavior ($P_{rad} \neq P_{tan}$) reflects the internal stresses necessary to maintain WH stability.
The behavior of these quantities is presented in Fig.~\ref{NEC}.
From Fig.~\ref{NEC}(a), both $\rho + P_{rad}$ and $\rho + P_{tan}$ exhibit regions of NEC violation near the throat, consistent with the need for exotic matter. The WEC is partially violated in the radial direction ($\rho + P_{rad} < 0$) but remains satisfied in the tangential direction.
Fig.~\ref{NEC}(b) shows the DEC behavior, near the throat, $\rho \gtrsim |P_{rad}|$ and $\rho \gtrsim |P_{tan}|$, indicating partial satisfaction of the DEC. At larger radial distances, DEC is fully satisfied, which suggests that exotic effects are confined to the vicinity of the throat, while the asymptotic spacetime behaves in accordance with standard physical conditions.
Consequently, the fuzzy WH model demonstrates localized NEC and WEC violations a standard feature of traversable WHs while preserving physical viability away from the throat, where the energy density dominates over pressure components.

\section{Complexity Factor}

In contrast to the mass profile, the complexity factor offers insight into the internal structural organization of the matter content, while a physical system is considered complex when multiple interacting components such as anisotropic stresses, energy density gradients, and pressure variations prevent a simple structural configuration. Herrera \cite{herrera2018new} introduced the notion of a complexity factor for spherically symmetric, static fluids in general theory of relativity, showing that a configuration is minimally complex (i.e., $Y_{TF}=0$) when the fluid possesses both homogeneous energy density and isotropic pressure. Interestingly, even when energy density is inhomogeneous and pressure stresses are anisotropic, a vanishing complexity factor may still arise if the contributions of these two effects cancel each other out \cite{andrade2023anisotropic,al2023complexity,naseer2024implications,naseer2024imprints,naseer2025wormhole,rehman2025interpretation,ashraf2026complexity}. 
The scalar $Y_{TF}$ effectively characterizes the system's complexity and mathematically is given by:
\begin{align}\label{a35a}
Y_{T F}=8 \pi \delta(r)-\frac{4 \pi}{r^3} \int_0^r \bar{r}^3 \rho^{\prime} d r
\end{align}
where $\delta(r)=P_{rad}-P_{tan}$. The vanishing complexity condition $Y_{T F}=0$ yields;
\begin{align}
\delta(r)=\frac{1}{2 r^3} \int_0^r {r}^3 \rho^{\prime} d {r}
\end{align}
It is worth emphasizing that this constraint is not limited to highly symmetric systems with isotropic as well as homogeneous behavior, but can also be realized in a broader class of gravitational configurations. In this setting, the assumption of vanishing complexity introduces a nonlocal equation of state, thereby providing the supplementary relation required to complete the Einstein field equations, whereas some similar approaches considered in investigations~\cite{casadio2019isotropization,contreras2021gravitational,arias2022anisotropic}. Now, we apply this formalism to study the complexity of the proposed WH solutions however for a finite throat radius $r_0$, Eq.~\eqref{a35a} becomes
\begin{align}\label{a35}
Y_{TF}&=8 \pi \delta(r)-\frac{4\pi}{r^{3}}\int_{r_0}^{r} r^{3}\rho'(r)\,dr
\end{align}
which reflects the combined influence of anisotropic pressure as well as spatial variations in the matter density. By adopting a constant redshift function and inserting the expressions for $\rho$, $P_{rad}$, and $P_{tan}$, we obtain
\begin{widetext}
\begin{align}\nonumber
&Y_{TF} =\frac{1}{(\lambda +4)^2 r^3}\left(3 (\lambda -2) (\lambda +4) r^{\frac{\lambda }{\lambda +4}} r_0^{\frac{4}{\lambda +4}}+4 \pi  \rho_s \left((\lambda +4)^2 r_0^3 \left(\left(\frac{r_0}{r_s}\right){}^{\frac{1}{\mathit{n}}}\right){}^{-3 \mathit{n}} \Gamma \left(3 \mathit{n}+1,\left(\frac{r_0}{r_s}\right){}^{\frac{1}{\mathit{n}}}\right)+24 (\lambda -2) \mathit{n} \right.\right.
\\\nonumber&r^{\frac{\lambda }{\lambda +4}} r_0^{\frac{2 (\lambda +6)}{\lambda +4}} \left(\left(\frac{r_0}{r_s}\right){}^{\frac{1}{\mathit{n}}}\right){}^{-\frac{2 (\lambda +6) \mathit{n}}{\lambda +4}} \Gamma \left(\frac{2 \mathit{n} (\lambda +6)}{\lambda +4},\left(\frac{r_0}{r_s}\right){}^{\frac{1}{\mathit{n}}}\right)+r^3 \left(-4 \left(2 \lambda ^2+7 \lambda -4\right) e^{-\left(\frac{r}{r_s}\right)^{\frac{1}{\mathit{n}}}}-(\lambda +4)^2 \left(\left(\frac{r}{r_s}\right)^{\frac{1}{\mathit{n}}}\right)^{-3 \mathit{n}} \right.
\\\nonumber&\left.\left.\left.\Gamma \left(3 \mathit{n}+1,\left(\frac{r}{r_s}\right)^{\frac{1}{\mathit{n}}}\right)-24 (\lambda -2) \mathit{n} \left(\left(\frac{r}{r_s}\right)^{\frac{1}{\mathit{n}}}\right)^{-\frac{2 (\lambda +6) \mathit{n}}{\lambda +4}} \Gamma \left(\frac{2 \mathit{n} (\lambda +6)}{\lambda +4},\left(\frac{r}{r_s}\right)^{\frac{1}{\mathit{n}}}\right)\right)\right)\right)
\\\label{a36}.
\end{align}
\end{widetext}

\begin{figure*}[htbp]
\centering
{{\includegraphics[height=2.0 in, width=3.2 in]{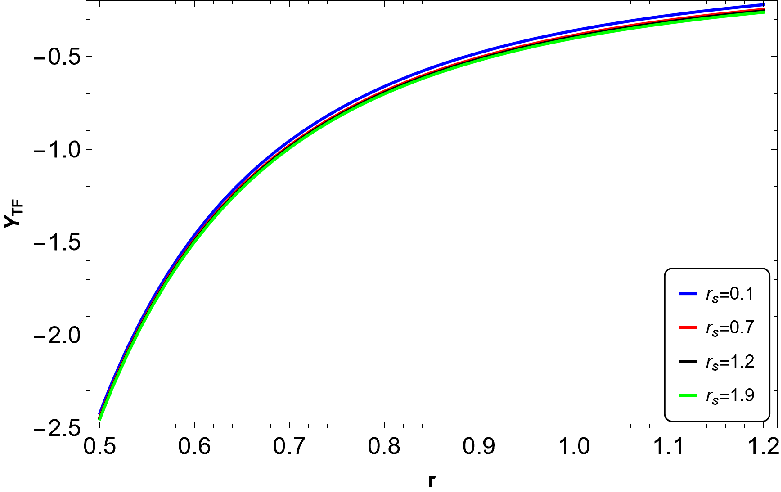}}}
\qquad
{{\includegraphics[height=2.0 in, width=3.2 in]{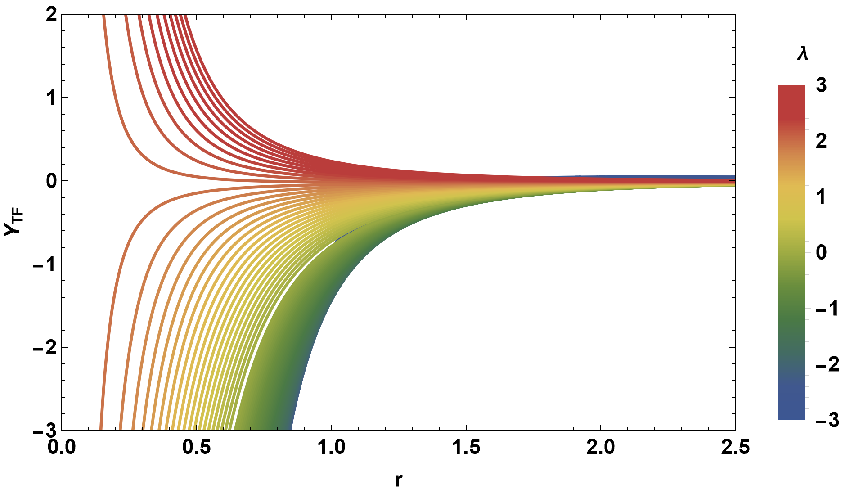}}}
\caption{Complexity factor $Y_{TF}$ versus radial coordinate $r$ for the same configurations.}
\label{massComplexity}
\end{figure*}
Plots~\ref{massComplexity} shows the behavior of the complexity factor $Y_{TF}$ with respect to the radial coordinate $r$ for the considered fuzzy WH configurations, while the left panel represents the radial evolution of $Y_{TF}$ for different values of the scale parameter $r_s$. It can be observed that $Y_{TF}$ remains negative throughout the plotted domain, indicating that the anisotropic contribution as well as density inhomogeneity play a significant role in the internal structure of the WH geometry. Near the throat $r_0=0.5$ region, the magnitude of $Y_{TF}$ is comparatively large, which suggests that the complexity of the system is more noticeable in the strong field region, however, as the radial distance increases, $Y_{TF}$ rises monotonically and approaches values close to zero and this decreasing magnitude of the complexity factor shows that the effect of local anisotropy and inhomogeneous matter distribution gradually weakens away from the throat. So, we first in left panel of Fig. \ref{massComplexity} examine the influence of the scale length $r_s$ on the complexity factor by fixing the Kalb Ramond parameter as $\lambda=1$, for this purpose, we take $r_0=0.5,~ \rho_s=0.01,~n=3.33,$ while the scale length is varied as $r_s=0.1,~0.7,~1.2,~1.9$. The variation of the parameter $r_s$ produces only a mild quantitative change in the profile of $Y_{TF}$ in this panel, as we observe for larger values of $r_s$, the curves shift slightly downward, showing a comparatively stronger negative contribution to the complexity factor, however all curves follow the same qualitative trend for different choices of parameter $r_s$.

Next, we study the dependence of complexity factor on the Kalb Ramond gravity parameter $\lambda$, in this case, the parameters are fixed as $r_0=0.5,~\rho_s=0.01,~r_s=0.1,~n=3.33,$ while the parameter $\lambda$ is varied continuously in the interval $-3 \leq \lambda \leq 3$.
The resulting behavior is shown in right panel of Fig.~\ref{massComplexity}. Consequently, the right panel displays the combined influence of the correction gravitational theory parameter $\lambda$ on the complexity factor, and the plot shows that the values of $Y_{TF}$ are highly sensitive to $\lambda$ in the region close to the WH throat $r_0=0.5$. For positive values of $\lambda$, the complexity factor tends to move toward positive branches near the inner region, whereas negative values of $\lambda$ drive the system toward more negative values of $Y_{TF}$, however, for sufficiently large radial distance, all profiles tend to merge and approach zero. This behavior indicates that the role of this parameter is dominant only near the throat, while its influence becomes negligible in the asymptotic region. Physically, these results suggest that the proposed fuzzy WH configuration is highly complex near the throat due to the combined effects of anisotropic pressure, inhomogeneous matter distribution, and modified gravity corrections. The gradual reduction of $Y_{TF}$ with increasing $r$ indicates that the WH spacetime becomes less complex away from the central region.

\section{Total Gravitational Energy, active gravitational mass and average pressure}\label{subsec:TGE}

\begin{figure*}[htbp]
\centering
{{\includegraphics[height=2.0 in, width=3.2 in]{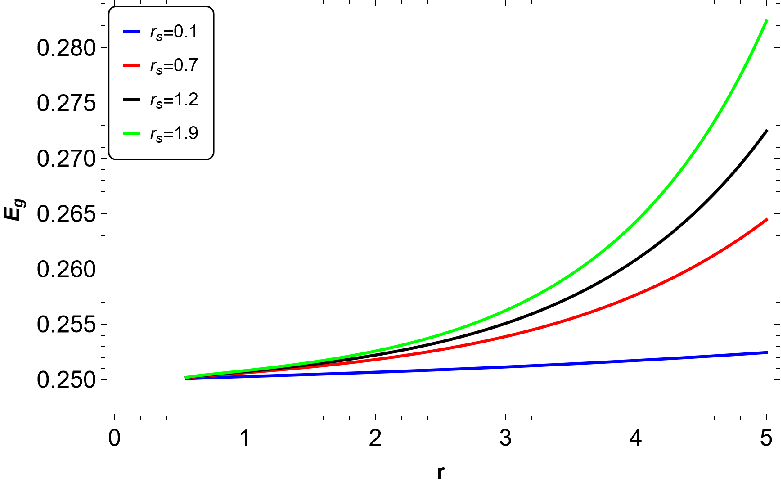}}}
\qquad
{{\includegraphics[height=2.0 in, width=3.2 in]{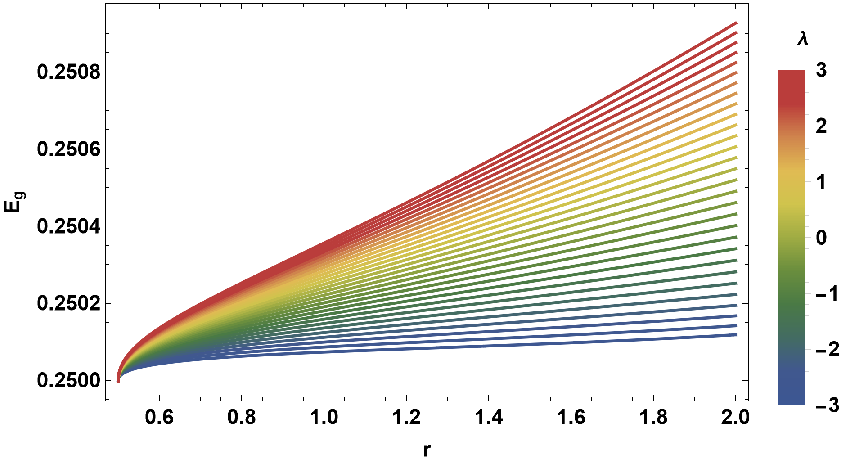}}}
\caption{In left panel behavior of the total gravitational energy $E_g$ versus the radial coordinate $r$ for different values of the scale length $r_s$, with fixed parameters $r_0=0.5$, $\rho_s=0.01$, $n=3.33$, and $\lambda=1$. In right panel behavior of the total gravitational energy $E_g$ versus $r$ with fixed parameters $r_0=0.5$, $\rho_s=0.02$, $r_s=0.7$, and $n=2$.}
\label{Egplots}
\end{figure*}

To further assess the physical acceptability of the constructed WH solution, we examine the behavior of the total gravitational energy. This quantity provides important insight into the combined influence of the matter distribution and the WH geometry outside the throat. From the preceding analysis, it has been found that the matter source supporting the proposed WH generally violates the radial NEC for most of the considered parameter space. Therefore, the corresponding source may be interpreted as exotic in nature. Since a configuration composed of ordinary baryonic matter usually possesses negative total gravitational energy, it is useful to explore whether the gravitational energy associated with the present galactic WH model exhibits attractive or repulsive behavior. For the dark-matter-inspired galactic WH under consideration, the total gravitational energy $E_g$ is defined as~\cite{lynden2007energy,nandi2009energetics}
\begin{align}\label{energy1}
E_g=M c^2-E_M .
\end{align}
Here, $M c^2$ denotes the total energy of the system and is written as
\begin{align}\label{energy2}
M c^2=\frac{r_0}{2}+\frac{1}{2}\int_{r_0}^{r}\left(T_t^t\right)^{(M)} r^2dr.
\end{align}
In this expression, the term $\frac{r_0}{2}$ corresponds to the effective mass contribution associated with the WH throat~\cite{nandi2009energetics}. Moreover, $E_M$ represents the total matter energy contribution, including rest energy, internal energy, kinetic energy, and other relevant forms of energy, and is given by
\begin{align}\label{energy3}&
E_M=\frac{1}{2} \int_{r_0}^r \sqrt{g_{r r}}  r^2\left(T_t^t\right)^{(M)} d r, \text { with } g_{r r}=\left(1-\frac{Y(r)}{r}\right)^{-1}
\end{align}
Consequently, total gravitational energy $E_g$ as given in eq. \eqref{energy1} yields the following expression;
\begin{align}\label{energy4}&
E_g=\frac{1}{2} \int_{r_0}^rr^2 d r\left[1-\sqrt{g_{r r}}\right]\left(T_t^t\right)^{(M)} +\frac{r_0}{2}
\end{align}
which implies that
\begin{align}\label{energy5}&
E_g=-\frac{1}{2}\int_{r_0}^{r}
\left[1-\left(1-\frac{Y(r)}{r}\right)^{-\frac{1}{2}}\right]
\rho r^{2}dr+\frac{r_0}{2}.
\end{align}
Accordingly, the total gravitational energy associated with the present model can be written as follows:
\begin{align}\label{energy6}&
E_g=\frac{r_0}{2}+\frac{1}{2}\int_{r_0}^{r}\left[\left(1-\frac{Y(r)}{r}\right)^{-1/2}-1\right]\rho\,r^2\,dr,
\end{align}
In the present study, the matter content is modeled through the Einasto dark matter density profile where $\rho_s$ corresponds to the dark matter density parameter, $h$ is the scale length, and $n$ is the Einasto index, which characterizes the shape of the Einasto dark matter profile. Moreover, $\lambda$ represents the Kalb Ramond gravity parameter. Since the obtained shape function contains incomplete Gamma functions, the above integral does not reduce to a simple closed analytical form. Therefore, It is quite difficult to obtain exact solutions of this integral due to its complicated form, so the total gravitational energy is evaluated numerically for different choices of the model parameters in order to explore its physical behavior. We first examine the influence of the scale length $r_s$ on the total gravitational energy by fixing the Kalb Ramond parameter as $\lambda=1$, for this purpose, we take $r_0=0.5,~ \rho_s=0.01,~\mathit{n}=3.33,$ while the scale length is varied as $r_s=0.1,~0.7,~1.2,~1.9.$ The corresponding behavior of $E_g$ as a function of the radial coordinate $r$ is displayed in left panel of Fig.~\ref{Egplots}, which shows that the total gravitational energy remains positive and increases monotonically with the radial coordinate for all chosen values of $r_s$. Near the throat, all curves start from approximately the same value, close to $E_g\approx 0.25$, and then gradually separate as $r$ increases which indicates that the contribution of the scale length becomes more pronounced away from the throat. A clear ordering of the curves is observed throughout the entire domain:
\[
r_s=1.9 > r_s=1.2 > r_s=0.7 > r_s=0.1.
\]
Thus, larger values of $r_s$ yield higher values of the total gravitational energy, in particular, the curve corresponding to $r_s=1.9$ grows most rapidly, whereas the curve for $r_s=0.1$ exhibits the slowest increase and remains the lowest among all cases. This behavior can be attributed to the role of the scale length in the Einasto profile, while a larger value of $r_s$ spreads the matter distribution over a wider radial region, thereby enhancing the cumulative contribution of the density profile in the integral expression of $E_g$, consequently, the total gravitational energy increases more significantly as $r_s$ becomes larger.
The smooth and monotonic nature of all curves also indicates that the WH geometry remains regular in the considered range, moreover, the absence of any crossing among the curves demonstrates that the effect of $r_s$ is systematic and consistent: increasing the scale length always strengthens the total gravitational energy contribution.

Next, we study the dependence of the total gravitational energy on the Kalb Ramond gravity parameter $\lambda$, in this case, the parameters are fixed as $r_0=0.5,~\rho_s=0.01,~r_s=0.1,~n=3.33,$
while the parameter $\lambda$ is varied continuously in the interval $-3 \leq \lambda \leq 3$.
The resulting behavior is shown in right panel of Fig.~\ref{Egplots}.
Right panel of Fig.~\ref{Egplots} reveals that the total gravitational energy again remains positive and increases monotonically with the radial coordinate for all values of $\lambda$. Very close to the throat, the curves are nearly indistinguishable, showing that the effect of $\lambda$ is initially weak. However, as the radial coordinate increases, the curves gradually fan out and become clearly separated, which indicates that the influence of the Kalb Ramond parameter becomes more significant at larger distances from the throat. It is evident from the figure that larger values of $\lambda$ correspond to larger values of the total gravitational energy. In other words, the upper curves are associated with positive values of $\lambda$, while the lower curves correspond to negative values of $\lambda$. Hence, the Kalb Ramond correction contributes positively to the energetic content of the WH configuration. The smooth spread of the family of curves further indicates that the parameter $\lambda$ modifies the geometry in a controlled manner without introducing any irregular behavior in the considered radial range.

Overall, the graphical analysis of both panels of Fig.~\ref{Egplots} demonstrate that the total gravitational energy is highly sensitive to the model parameters, whereas the scale length $r_s$ governs the radial spread of the Einasto dark matter distribution and significantly enhances the magnitude of $E_g$ for larger values. On the other hand, the Kalb Ramond gravity parameter $\lambda$ also increases the total gravitational energy and produces a systematic upward shift in the family of curves, hence, the behavior of $E_g$ confirms that both the matter distribution and the modified gravity correction play a crucial role in determining the energetic structure of the WH geometry. According to Misner's interpretation, the total gravitational energy behaves attractively when $E_g<0$, whereas it exhibits a repulsive character when $E_g>0$. The graphical results displayed in Fig.~\ref{Egplots} show that $E_g$ remains positive in both cases which indicates that the total gravitational energy produces a repulsive effect near the WH throat. Such repulsive behavior is physically relevant, as it can help counterbalance gravitational collapse as well as thereby support the existence of viable galactic WH configurations in the scenario of Kalb Ramond gravity.

Let us present a unified analysis of an active gravitational mass and average pressure associated with fuzzy WH configurations. These two aforementioned physical quantities complement each other in understanding how matter distribution, anisotropic stresses, and energy density variations collectively govern the internal structural behavior of the system. However, together these quantities provide a comparative picture of how the fuzzy WH evolves from its throat to the outer regions. The active gravitational mass $M_a$ quantifies the amount of matter enclosed between the WH throat radius $r_0$ and any radial position $r$, whereas for a given fuzzy WH density profile, it is defined as
\begin{align}\label{a27}
M_{a}&=4\pi\int_{r_0}^{r}\rho(r)\, r^{2}\,dr.
\end{align}
and after evaluation one conceive the relation for active gravitational mass as follows:
\begin{widetext}
\begin{align}\label{a28}
M_{a}=4 \pi  \rho_s \left(\mathit{n} r_0^3 \left(\left(\frac{r_0}{r_s}\right){}^{\frac{1}{\mathit{n}}}\right){}^{-3 \mathit{n}} \Gamma \left(3 \mathit{n},\left(\frac{r_0}{r_s}\right){}^{\frac{1}{\mathit{n}}}\right)-\mathit{n} r^3 \left(\left(\frac{r}{r_s}\right)^{\frac{1}{\mathit{n}}}\right)^{-3 \mathit{n}} \Gamma \left(3 \mathit{n},\left(\frac{r}{r_s}\right)^{\frac{1}{\mathit{n}}}\right)\right).
\end{align}
\end{widetext}
To assess the pressure behavior in all spatial directions, the average pressure for the anisotropic fluid is defined as
\begin{equation}
P_{aver} =\frac{\big(P_{\mathrm{rad}} + 2P_{\mathrm{tan}}\big)}{3},
\end{equation}
\begin{figure*}[htbp]
\centering
{{\includegraphics[height=2.0 in, width=3.2 in]{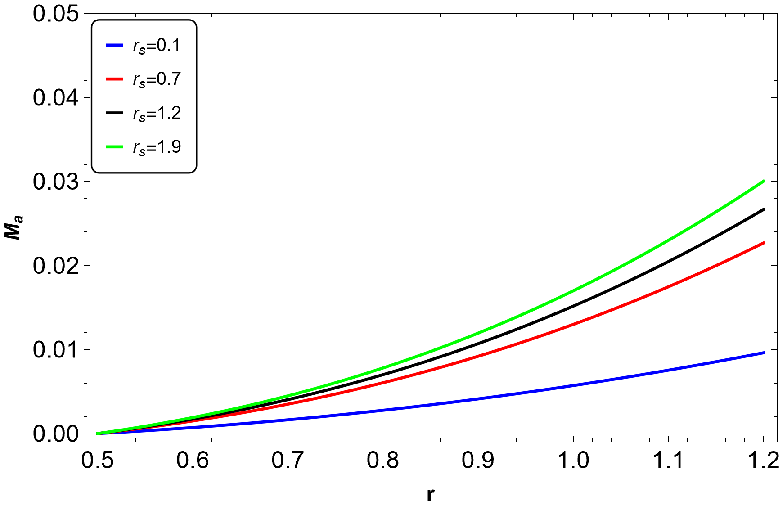}}}
\qquad
{{\includegraphics[height=2.0 in, width=3.2 in]{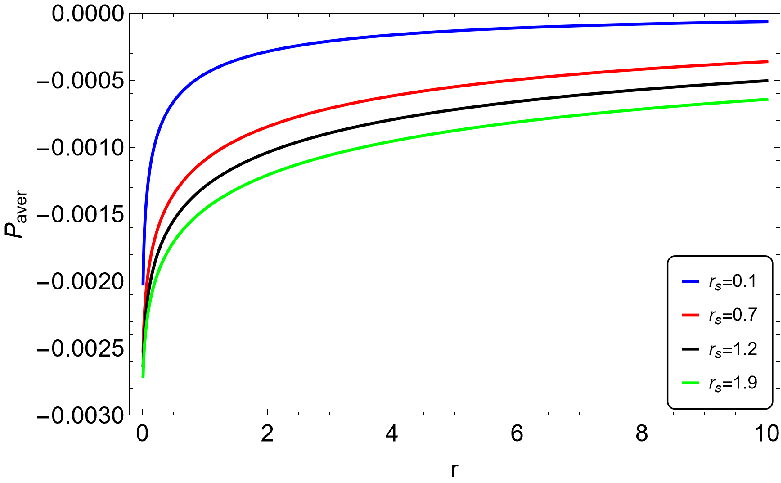}}}
\caption{For panel (a) active gravitational mass $M_{a}$ for fuzzy WHs with $r_0=0.5$, shown for the Einasto index $n=3.33$ and panel (b) average pressure versus radial coordinate $r$ for the same configurations.}
\label{masspaver}
\end{figure*}
Fig.~\ref{masspaver}(a) displays the radial evolution of $M_a$ for different values of the Einasto index $n$ and an increase in $n$ leads to a stronger rise in $M_a$, demonstrating that the gravitational mass grows more rapidly as the density distribution becomes sharper. A narrow region near the throat shows slightly negative values of $M_a$, signalling the presence of exotic matter required to sustain the traversable WH geometry; however, as $r$ increases, the WH mass becomes positive and monotonically increasing.
For a physically plausible configuration, $P_{aver}$ must remain finite and positive while decreasing outward, but in our considered case it can be see as negative which indicating existance of exotic matter that is necessary for sustaining traversable WH as illustrated in Fig.~\ref{masspaver}.

\section{Equilibrium Analysis through the TOV Equation and Pressure Anisotropy}\label{Stability}

\begin{figure*}[htbp]
\centering
{{\includegraphics[height=2.0 in, width=3.2 in]{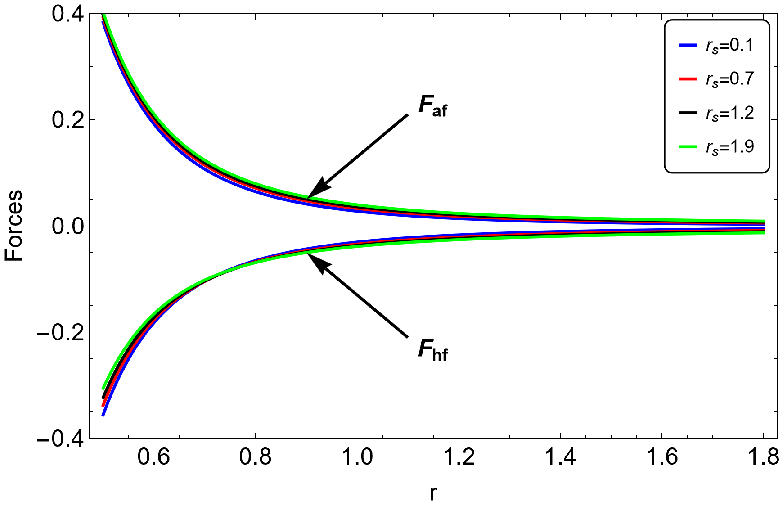}}}
\qquad
{{\includegraphics[height=2.0 in, width=3.2 in]{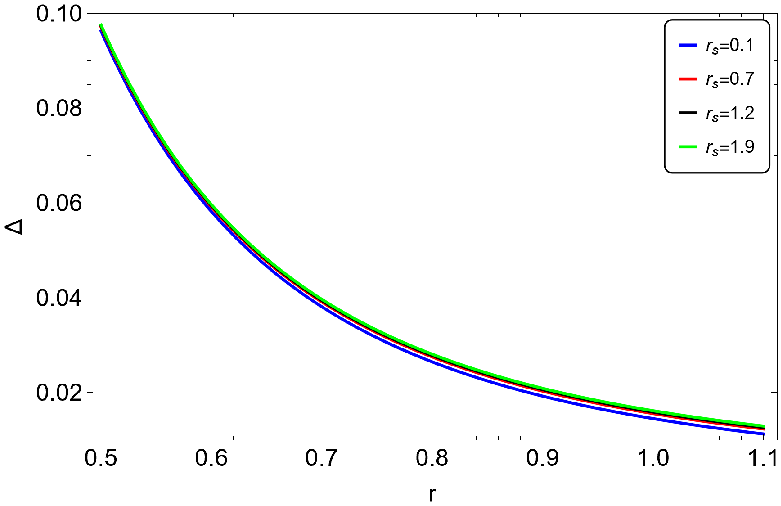}}}
\caption{Pictorial representation of equilibrium forces panel (a), anisotropy parameter panel (b) and average pressure panel (c) at $r_0=0.5$ and $n=3.33$ and varying Einasto-density profile model parameters.}\label{TOVAnAP}
\end{figure*}

The equilibrium properties of the constructed WH configurations can be
examined through the generalized TOV equation,
which describes the balance among the effective force contributions generated
by the anisotropic matter distribution. In the present model, the energy
density and pressure components are obtained from the Kalb Ramond modified
gravitational field equations together with the Einasto dark matter profile, consequently, the equilibrium behavior reflects the combined influence of the
Einasto matter distribution and the Kalb Ramond gravitational parameter
$\lambda$.
For an anisotropic fluid configuration, the generalized TOV equation may be
written as
\begin{equation}\label{GTOV}
\frac{dP_{\mathrm{rad}}}{dr} = -\frac{A_{1}'(r)}{2}\big(\rho + P_{\mathrm{rad}}\big) - \frac{2}{r}\big(P_{\mathrm{rad}} - P_{\mathrm{tan}}\big),
\end{equation}
In above mention TOV equation \eqref{GTOV} $A_{1}'(r)$ denotes the derivative of metric potential with respect to radial coordinate $r$, and this expression can be decomposed into three force components:
\begin{description}
  \item[Gravitational Force:] $F_{\mathrm{gf}} = -\frac{A_{1}'(r)}{2}\,(\rho+P_{\mathrm{rad}})$
  \item[Anisotropic Force:] $F_{\mathrm{af}} = -\frac{2}{r}\,(P_{\mathrm{rad}}-P_{\mathrm{tan}})$
  \item[Hydrostatic Force:] $F_{\mathrm{hf}} = -\frac{dP_{\mathrm{rad}}}{dr}$
\end{description}
representing gravitational, anisotropic, and hydrostatic forces, respectively. The equilibrium condition requires
\begin{equation}
F_{\mathrm{gf}} + F_{\mathrm{af}} + F_{\mathrm{hf}} = 0.
\end{equation}
Since the present WH configuration is constructed under the
zero tidal force assumption with a constant redshift function is constant and its radial derivative vanishes, consequently, the gravitational force contribution in the
generalized TOV equation becomes zero, and the equilibrium condition reduces
to a balance between the hydrostatic and anisotropic forces,
\begin{equation}
F_{\mathrm{af}}+F_{\mathrm{hf}}=0.
\end{equation}
A physically viable model requires the radial pressure to be smooth and monotonic behavior, in this case study it ensuring existence of exotic matter.
The pressure anisotropy is defined as
\begin{equation}
\Delta(r)=P_{\mathrm{tan}}-P_{\mathrm{rad}}.
\end{equation}
There three basic possibilities for the value of $\Delta$ as;
A positive value, $\Delta>0$, corresponds to an outward directed
anisotropic contribution, whereas $\Delta<0$ indicates an inward directed
anisotropic force. The isotropic limit is recovered for $\Delta=0$.
From Fig. \ref{TOVAnAP}, we can observe that:
\begin{description}
  \item[a)] Fig.~\ref{TOVAnAP}(a) demonstrates that the forces $F_{\mathrm{af}}$ and $F_{\mathrm{hf}}$ counterbalance each other, indicating the possible existence of a static, spherically symmetric fuzzy WH geometry. This stability is consistent with the parameter choices previously shown to satisfy energy conditions.
  \item[b)] Fig.~\ref{TOVAnAP}(b) shows that $\Delta(r)$ remains positive and the Einasto-density profile model parameters, signifying the presence of a repulsive anisotropic force that supports the WH structure against collapse.
\end{description}
The graphical behavior of the hydrostatic and anisotropic force components
shows that they counterbalance each other throughout the considered radial
domain, consistently with the equilibrium condition obtained for the
constant redshift configuration. Moreover, the positive pressure anisotropy
provides an outward directed contribution that assists in maintaining the
WH throat against inward contraction. These results indicate that the
constructed Einasto supported WH configurations can maintain static
mechanical equilibrium within the considered parameter domain of
Kalb Ramond gravity.

\section{Optical Properties of Einasto Supported WHs in Kalb Ramond Gravity}\label{sec:light_echo}

In this section, we investigate the optical properties of the constructed
Einasto supported WH configuration within the scenario of
Kalb Ramond gravity, in particular, we analyze null geodesics and examine
two important observables, namely the light deflection angle and the echo
time delay. The spacetime geometry is described by the static as well as
spherically symmetric Morris-Thorne metric with a constant redshift
function, whereas the shape function $Y(r)$ is obtained by combining the
Einasto dark matter density profile with the Kalb Ramond modified field
equations. Consequently, the optical behavior of the system is determined
by the geometrical structure encoded in $Y(r)$ as well as depends on the
characteristic Einasto parameters together with the Kalb Ramond
gravitational parameter $\lambda$. For simplicity, we restrict the photon
motion to the equatorial plane, $\theta=\pi/2$. Since the spacetime is
static and spherically symmetric, the time coordinate $t$ and the
azimuthal coordinate $\phi$ are cyclic, leading to the conservation of
photon energy and angular momentum along the null geodesics \cite{ditta2026impact}.
Thus, for an affine parameter \(\tilde{\lambda}\), the conserved quantities are
\begin{equation}
E\equiv -g_{tt}\dot{t}=\dot{t},
\qquad
L\equiv g_{\phi\phi}\dot{\phi}=r^{2}\dot{\phi},
\end{equation}
where an overdot denotes differentiation with respect to \(\tilde{\lambda}\). Using the null condition
\begin{equation}
g_{\eta\aleph}\dot{x}^{\eta}\dot{x}^{\aleph}=0,
\end{equation}
we obtain the following expression
\begin{equation}
-\dot{t}^{2}+\frac{\dot{r}^{2}}{1-\dfrac{Y(r)}{r}}+r^{2}\dot{\phi}^{2}=0.
\end{equation}
here, first term represent time contribution, second term radial motion and third term angular motion.
After substituting the conserved quantities, the radial equation becomes
\begin{equation}
\dot{r}^{2}=\left(1-\frac{Y(r)}{r}\right)
\left(E^{2}-\frac{L^{2}}{r^{2}}\right).
\end{equation}
This expression shows that the radial motion of photon depends on geometry and conserved quantities. Introducing the impact parameter which is an important quantity in light deflection and photon sphere analysis, and written as
\begin{equation}
b_{\rm imp}=\frac{L}{E},
\end{equation}
the above expression may be rewritten as
\begin{equation}
\dot{r}^{2}=E^{2}\left(1-\frac{Y(r)}{r}\right)
\left(1-\frac{b_{\rm imp}^{2}}{r^{2}}\right).
\end{equation}

The corresponding effective potential for photon motion can therefore be written as
\begin{equation}
V_{\mathrm{eff}}(r)=
\left(1-\frac{Y(r)}{r}\right)\frac{L^{2}}{r^{2}}.
\end{equation}
The unstable circular photon orbit, or photon sphere, is determined from the extremum condition
\begin{equation}
\frac{d}{dr}\left[\frac{\left(1-\dfrac{Y(r)}{r}\right)}{r^{2}}\right]=0,
\end{equation}
which, reduces to
\begin{equation}
r\,Y'(r)-3Y(r)+2r=0.
\end{equation}
Its largest positive root gives the photon sphere radius \(r_{\mathrm{ph}}\) as shown in Fig. (6), whereas for a circular photon orbit at $r=r_{\rm ph}$, one must have $\dot r=0$. Hence, the impact parameter takes its critical value at the photon sphere. Using the null geodesic relation together with the metric coefficient of the radial sector, so the corresponding critical impact parameter is then
\begin{equation}
b_{\rm imp(\rm crit)}^{2}
=\frac{r_{\rm ph}^{2}}{\left(1-\dfrac{Y(r_{\rm ph})}{r_{\rm ph}}\right)},
\end{equation}
\text{therefore}
\begin{equation}
b_{\rm imp(\rm crit)}=\frac{r_{\mathrm{ph}}}{\sqrt{1-\dfrac{Y(r_{\mathrm{ph}})}{r_{\mathrm{ph}}}}}.
\end{equation}

\subsection{Analyzing Deflection Angle}\label{def_angle}

\begin{figure*}[htbp]
\centering
{{\includegraphics[height=2.0 in, width=3.0 in]{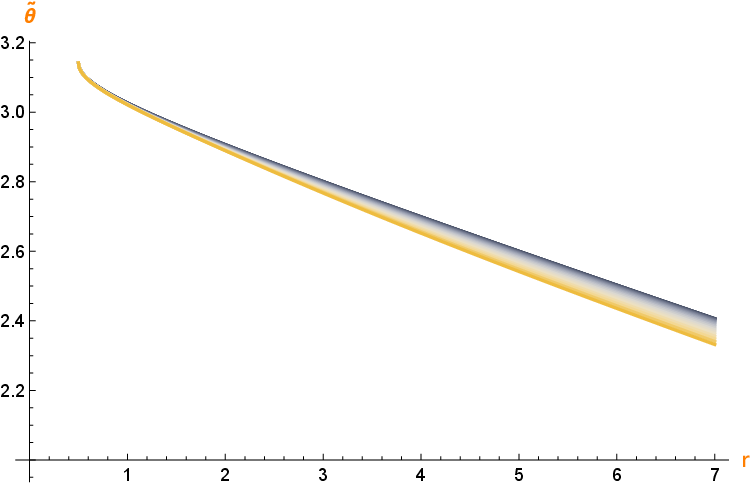}}}
{{\includegraphics[height=2.0 in, width=0.2 in]{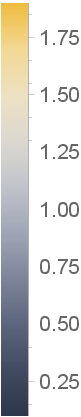}}}
\qquad
{{\includegraphics[height=2.0 in, width=3.0 in]{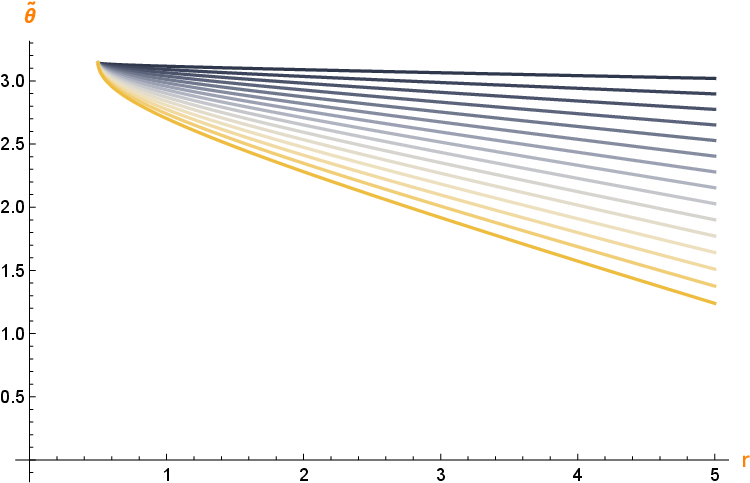}}}
{{\includegraphics[height=2.0 in, width=0.2 in]{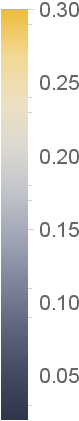}}}
\caption{Variation of the light-deflection angle $\tilde{\theta}$ with the Einasto scale radius $r_s$ (left panel) and the impact parameter (right panel), keeping the other model parameters fixed.}
\label{deflection_angle}
\end{figure*}

To determine the bending of light in the present Einasto WH geometry, we utilize following expressions
\begin{equation}
\frac{dr}{d\phi}=\frac{dr/d\tilde{\lambda}}{d\phi/d\tilde{\lambda}}=\frac{\dot r}{\dot\phi},~~\text{where}~~\dot\phi=\frac{L}{r^2}=\frac{b_{\rm imp} E}{r^2}
\end{equation}
which yields
\begin{equation}
\frac{dr}{d\phi}=\pm r^{2}
\sqrt{\left(1-\frac{Y(r)}{r}\right)
\left(\frac{1}{b_{\rm imp}^{2}}-\frac{1}{r^{2}}\right)}.
\end{equation}
Here, the plus sign shows the outward motion and the minus symbol represents the inward motion.
The distance of closest approach \(r_{\min}\) is obtained from the turning point condition
\begin{equation}
\left(1-\frac{Y(r)}{r}\right)
\left(\frac{1}{b_{\rm imp}^{2}}-\frac{1}{r^{2}}\right)=0.
\end{equation}
Hence, the total deflection angle of photons that come from infinity and return to infinity is
\begin{equation}
\tilde{\theta}=2\int_{r_{\min}}^{\infty}
\frac{dr}
{r^{2}
\sqrt{\left(1-\dfrac{Y(r)}{r}\right)
\left(\dfrac{1}{b_{\rm imp}^{2}}-\dfrac{1}{r^{2}}\right)}}
-\pi.
\label{eq:deflection_Einasto}
\end{equation}
The behavior of the deflection angle is displayed in the left and right panels of Fig. \ref{deflection_angle}. The left panel shows the variation of \(\tilde{\theta}\) for different values of the Einasto scale radius \(r_{s}\), whereas the right panel illustrates its dependence on the impact parameter \(b_{\rm imp}\). The graphical behavior indicates that the bending of light is strongest near the throat and gradually decreases as the radial coordinate increases, which is consistent with the fact that the spacetime curvature is more pronounced in the inner region of the WH and weakens asymptotically. Moreover, varying \(r_{s}\) changes the concentration of the Einasto matter distribution and, therefore, modifies the optical response of the WH geometry. In particular, larger values of \(r_{s}\) alter the profile of the shape function $Y(r)$ and shift the deflection curves, showing that the scale radius plays an important role in controlling the lensing strength of the system. The right panel further shows that the deflection angle is sensitive to the impact parameter: light rays with smaller \(b_{\rm imp}\) pass closer to the throat and experience stronger bending, whereas larger values of \(b_{\rm imp}\) correspond to weaker deflection. Therefore, Fig.~\ref{deflection_angle} confirms that the constructed Einasto WH possesses a nontrivial lensing signature and that its optical properties can be regulated by the model parameters.

\subsection{Probe of Echo Time}\label{Echo_time}

\begin{figure*}[htbp]
\centering
{\includegraphics[height=2.0 in, width=3.2 in]{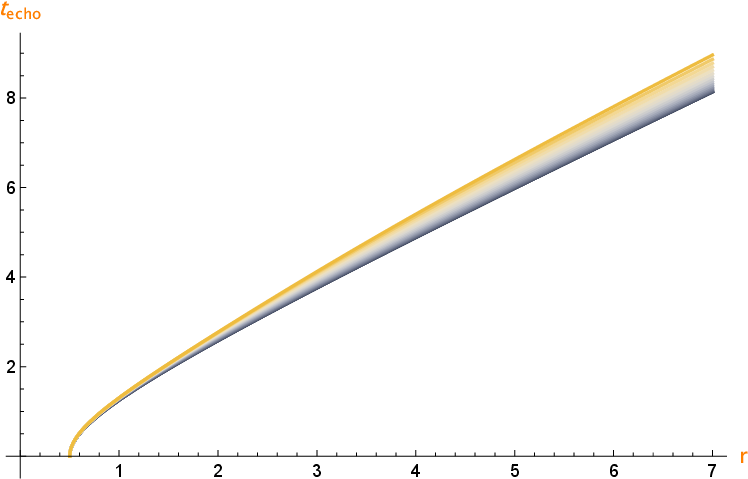}}
{{\includegraphics[height=2.0 in, width=0.3in]{figad-1.eps}}}
\caption{Behavior of the echo time $t_{\rm echo}$ for different values of
the Einasto scale radius $r_s$, with the remaining model parameters kept
fixed.}
\label{echo_time}
\end{figure*}

Another important observable associated with horizonless compact geometries is the echo time delay. In traversable WH physics, this quantity estimates the round trip propagation time of a signal between the throat region and the outer potential barrier. For the present spacetime, the tortoise coordinate is defined by
\begin{equation}
\frac{dr_{*}}{dr}=\frac{1}{\sqrt{1-\dfrac{Y(r)}{r}}},
\end{equation}
and the geometric estimate of the echo time is given by
\begin{equation}
t_{\mathrm{echo}}\approx 2\int_{r_0}^{r_{\mathrm{ph}}}
\frac{dr}{\sqrt{1-\dfrac{Y(r)}{r}}}.
\label{eq:echo_Einasto}
\end{equation}
Here, \(r_0\) is the throat radius and \(r_{\mathrm{ph}}\) is the photon sphere radius obtained from the above extremum condition.
The behavior of \(t_{\mathrm{echo}}\) for different values of the Einasto scale radius \(r_{s}\) is presented in Fig.~\ref{echo_time}. The figure shows that the echo time remains finite throughout the considered parameter range, which supports the absence of any horizon-like trapping and is fully compatible with the traversable nature of the obtained WH geometry. It is also evident that changing Einasto scale radius \(r_{s}\) modifies the round trip travel time, since the scale radius directly affects the shape function $Y(r)$ and hence the effective radial extension of the throat region. As a consequence, different values of \(r_{s}\) lead to different echo delay profiles. This demonstrates that the temporal response of the Einasto WH is not universal but depends sensitively on the dark matter distribution encoded in the model. Therefore, Fig.~\ref{echo_time} indicates that the echo time provides an additional observational channel for studding the geometric structure of the WH and the influence of the Einasto matter parameters.
The combined analysis of the light deflection angle and echo time demonstrates
that the optical and temporal characteristics of the present WH geometry
are directly governed by the shape function $Y(r)$. Since $Y(r)$ is determined
by the Einasto dark matter distribution together with the Kalb Ramond
gravitational contribution, these observables inherit a dependence on the
characteristic Einasto parameters and the Kalb Ramond parameter $\lambda$,
in particular, the Einasto scale radius $r_s$ modifies the radial distribution
of matter and consequently alters the bending of light as well as the propagation
delay of signals in the vicinity of the WH throat. The light deflection
angle characterizes the modification of photon trajectories caused by the
WH geometry, whereas the echo time provides complementary information
about signal propagation in the corresponding horizonless spacetime.
Consequently, these optical observables provide useful probes of the interplay
between the Einasto matter distribution and the Kalb Ramond gravitational
sector in determining the observational characteristics of the constructed
WH configuration.

\subsection{Estimation of Exotic Matter through Volume Integral Quantifier}
\label{subsec:VIQ}

\begin{figure*}[htbp]
\centering
{{\includegraphics[height=2.6 in, width=3.0 in]{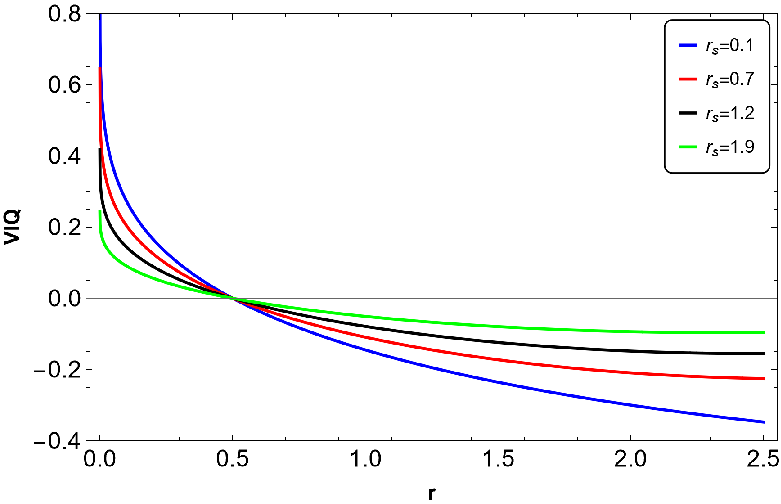}}}
\qquad
{{\includegraphics[height=2.6 in, width=3.6 in]{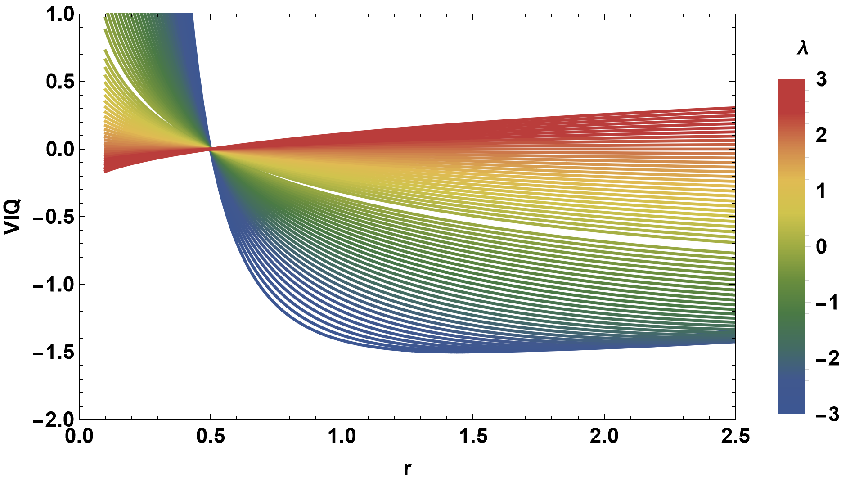}}}
\caption{Pictorial representation of VIQ vs $r$. Left panel shows behavior of the VIQ for different values of the Einasto scale radius \(r_s\) at fixed \(\lambda=1\) and right panel shows for different values of gravity correction parameter \(\lambda\) and $r_s=0.1$, while all other parametric values are fixed as \(r_0=0.5\), \(\rho_s=0.01\), and \(n=3.33\).}
\label{BothVIQ}
\end{figure*}

To estimate the total amount of such exotic matter, we use VIQ purposed Visser et al. in \cite{visser2003traversable} which measures the integral contribution of the NEC violating matter over a finite spatial region and is defined as

\begin{align}\label{viq1}
VIQ= \oint \big(\rho+P_{rad}\big) dV,
\end{align}
where \( dV = r^2 dr\, d\Omega \) is the proper volume element and \( d\Omega \) is the element of solid angle.
Since
$\oint dV = 2\int_{r_0}^{\infty} dV = 8\pi\int_{r_0}^{\infty}  r^2 dr,$
the entire NEC breaching is described in subsequent form:
$VIQ = 8\pi \int_{r_0}^{\infty} r^2 \big(\rho+P_{rad}\big) dr.$
We establish an upper bound to confine the scope to a substantially significant region \( R_1 \geq r_0 \), yielding:
\begin{equation}
VIQ(r)=8\pi \int_{r_{0}}^{R_1}\left(\rho+P_{rad}\right) r^{2}\,d r,
\label{VIQ-def}
\end{equation}
where \(R_1\) is the upper integration limit and \(r\) is the radial coordinate, while term \(\rho+P_{rad}\) represents the radial NEC. Therefore, the negative behavior of VIQ indicates the presence of exotic matter, whereas its tendency toward zero suggests that the total amount of exotic matter can be minimized.
For the present WH model, we obtain the VIQ by substituting the Einasto dark matter density profile and the corresponding radial pressure component into Eq.~\eqref{VIQ-def}. The resulting expression depends on the throat radius \(r_0\), the central density parameter \(\rho_s\), the Einasto index $n$, the scale radius \(r_s\), and the gravity correction parameter \(\lambda\), whereas throughout the graphical analysis, we take $r_0=0.5,~\rho_s=0.01,~n=3.33$.
The effects of the correction parameter \(\lambda\) and the Einasto scale radius \(r_s\) are then investigated separately, whereas right panel of Fig.~\ref{BothVIQ} shows the behavior of VIQ vs \(r\) for different values of the correction parameter \(\lambda\), while fixing the scale radius as \(r_s=0.1\), while the parameter \(\lambda\) is varied in the interval \(-3<\lambda<3\), excluding \(\lambda=0\), since the analytical expression of the VIQ contains inverse powers of \(\lambda\), whereas one can clearly observe that the curves exhibit a strong dependence on \(\lambda\). In particular, the VIQ becomes negative for a wide range of the radial coordinate, which confirms the existence of exotic matter in the WH configuration, whuile this negative contribution is directly related to the violation of the radial NEC, which is necessary to keep the WH throat open.
An important feature of right panel of Fig.~\ref{BothVIQ} is that the magnitude of VIQ decreases as the radial distance increases, while curves gradually approach the zero line in the outer region, showing that the exotic matter contribution becomes very small away from the throat. This behavior indicates that the exotic matter is mainly concentrated near the throat and does not dominate the entire spacetime, hence, the constructed fuzzy WH model requires only a localized distribution of NEC violating matter. Moreover, different choices of \(\lambda\) modify the strength of the exotic matter contribution, while negative values of \(\lambda\) generally produce a stronger negative behavior, whereas positive values move the VIQ closer to zero, consequently, the correction parameter \(\lambda\) provides a useful control on the amount of exotic matter required to support the WH geometry.

To further examine the role of the Einasto scale radius, left panel of Fig.~\ref{BothVIQ} presents the variation of VIQ with respect to \(r\) for different values of \(r_s\), while fixing the correction parameter at \(\lambda=1\). The left panel of this plot shows that the VIQ initially takes positive values in the inner region and then decreases monotonically as the radial coordinate increases, however after crossing the zero line near the throat region, the VIQ becomes negative, indicating the appearance of exotic matter required for the stability and maintenance of the WH structure.
It is clear from left panel of Fig.~\ref{BothVIQ} that the scale radius \(r_s\) significantly affects the magnitude of the VIQ, whereas for smaller values of \(r_s\), such as \(r_s=0.1\), the curve decreases more rapidly and attains larger negative values at higher radial distances. This means that smaller scale radii enhance the effective exotic matter contribution, while on the other hand, for larger values of \(r_s\), the curves remain closer to the zero line, which indicates that the magnitude of exotic matter is reduced. Therefore, at fixed \(\lambda=1\), increasing the Einasto scale radius weakens the negative contribution of the VIQ and consequently reduces the amount of exotic matter required in the model, whereas from both left and right panels of Fig.~\ref{BothVIQ}, we conclude that the VIQ analysis supports the physical viability of the present fuzzy WH solution. The negative values of VIQ confirm the presence of exotic matter, while the approach of the curves toward zero shows that the required amount of exotic matter can be made finite as wll as relatively small. Furthermore, the parameters \(\lambda\) and \(r_s\) play an important role in regulating the distribution and strength of the exotic matter, hence, the present model admits a WH configuration in which the exotic matter is mainly localized near the throat and can be controlled through suitable choices of the model parameters.

\section{Conclusion}

In this work, we investigated static and spherically symmetric traversable
WH configurations within the scenario of Kalb Ramond gravity, where
the gravitational sector is influenced by a spontaneously Lorentz symmetry breaking
Kalb Ramond background. The matter distribution supporting the WH
geometry modeled through the Einasto dark matter density profile,
however by combining the corresponding density distribution with the modified
gravitational field equations, we obtained an analytical expression for the
WH shape function as well as examined its geometrical and physical properties.
The principal results of the present analysis may be summarized as follows:
Firstly, the derived shape function satisfies the throat condition
$Y(r_0)=r_0$ and exhibits the required flare-out behavior within the
considered parameter domain, while the ratio $Y(r)/r$ decreases outside the throat,
confirming the geometrical viability of the resulting traversable WH
configuration. The corresponding two and three dimensional embedding
diagrams further illustrate the characteristic two sheeted WH structure
and the influence of the model parameters on the spatial curvature.
Secondly; the energy density remains positive in the considered region, whereas the
radial pressure becomes negative close to the WH throat, however the analysis
of the energy conditions shows that the radial null energy condition is
violated in the vicinity of the throat, indicating the presence of exotic
matter required to sustain the traversable geometry. At larger radial
distances, the exotic contribution becomes progressively weaker, suggesting
that the violation can remain confined to a restricted neighborhood of the
throat.
Thirdly; the complexity factor exhibits its strongest behavior in the vicinity of the
throat as well as gradually approaches zero with increasing radial distance, whereas this
behavior shows that the combined contribution of density inhomogeneity as well as
pressure anisotropy is predominantly localized in the strong field region,
while the outer spacetime becomes structurally less complex.
Fourthly; the total gravitational energy remains positive for the parameter choices
considered in the graphical analysis, indicating a repulsive gravitational
character that may contribute to maintaining the WH throat against
collapse. The active gravitational mass and average pressure further provide
complementary information about the distribution of matter and stresses
throughout the WH geometry.
Fifthly; the equilibrium analysis based on the generalized
TOV equation shows that, under the constant redshift
assumption, the gravitational force contribution vanishes and the mechanical
equilibrium is governed by the balance between the hydrostatic and anisotropic
forces, whereas the positive pressure anisotropy in the considered region provides an
outward directed contribution that assists in supporting the WH configuration.
Sixthly; the optical characteristics of the geometry investigated through
the photon sphere radius, critical impact parameter, light deflection angle,
and echo time, whereas these quantities are governed by the shape function and,
therefore, inherit the influence of both the Einasto dark matter distribution
as well as the Kalb Ramond gravitational parameter $\lambda$. The corresponding
behavior demonstrates that the obtained WH geometry possesses
nontrivial optical and temporal signatures. The volume integral quantifier confirms
that the null energy condition violating matter is mainly concentrated near the throat.
Its tendency toward smaller magnitude away from the inner region indicates that the total exotic
matter contribution can remain finite and may be reduced through suitable
choices of the Einasto and Kalb Ramond model parameters.
Overall, the present analysis demonstrates that the interplay between the
Einasto dark matter distribution and the Kalb Ramond gravitational sector
can support physically consistent traversable WH geometries over
appropriate parameter ranges. The Kalb Ramond parameter $\lambda$ modifies
the underlying geometry and associated physical observables, while the
Einasto parameters regulate the radial distribution of the matter source.
The resulting scenario therefore provides a useful setting for investigating
the geometrical, energetic, equilibrium, as well as optical properties of
dark matter supported WHs in Lorentz symmetry breaking gravitational
scenarios.

\end{document}